\documentstyle[emulateapj,psfig]{article} 
\slugcomment{final version, to appear in ApJ}  
\lefthead{ZHANG \& HARDING}  
\righthead{FULL POLAR CAP CASCADE SCENARIO}  
  
\begin{document}  
\title{Full polar cap cascade scenario: $\gamma$-ray and X-ray luminosities  
 from spin-powered pulsars}  
\author{Bing Zhang\altaffilmark{1} and Alice K. Harding}  
\affil {Laboratory of High Energy Astrophysics, NASA/Goddard Space 
 Flight Center, Greenbelt, MD 20771\\ 
 bzhang@twinkie.gsfc.nasa.gov, harding@twinkie.gsfc.nasa.gov} 
\altaffiltext{1} {National Research Council Research Associate Fellow.} 
 
\date{3-23-1999}  
  
\begin{abstract}

Canonical polar cap cascade models involve curvature radiation
(CR) or inverse Compton scattering (ICS) of the primary particles and
synchrotron radiation (SR) of the higher generation pairs. Here we
modify such a cascade picture to include the ICS of the higher
generation pairs. In such a ``full-cascade'' scenario, not only the
perpendicular portion of the energy of the pairs goes to high energy
radiation via SR, but the parallel portion of the energy of the pairs can  
also contribute to high energy emission via ICS with the soft thermal
photons from either the full neutron star
surface or the hot polar cap. The efficiency of converting particles'
kinetic energy to radiation by ICS is very high if the scatterings occur
in the ``resonant'' regime. As a result, almost 100\% of the energy input
from the pulsar inner accelerators could be converted to high energy
emission. An important output of such a scenario is that the soft  
tail of the ICS spectrum can naturally result in a non-thermal  
X-ray component which can contribute to the luminosities observed by
ROSAT and ASCA.  
 
Here we present an analytic description of such a full polar cap 
cascade scenario using the recursion relationships between adjacent 
generations following the approach first proposed by Lu et al. (1994), 
but develop it to be able to delineate the complex full-cascade 
process. The acceleration model we adopted is the space-charge-limited
flow model proposed by Harding \& Muslimov (1998). We present the 
theoretical predictions of the $\gamma$-ray luminosities, the thermal 
and non-thermal X-ray luminosities for the known spin-powered X-ray 
pulsars (8 of them are also $\gamma$-ray pulsars), and compare them 
with the observations from CGRO, ROSAT and ASCA. We estimate the
non-thermal X-ray luminosity by including all the possible ICS branches
contributing to a certain energy band, and estimate both the full
surface and hot polar cap thermal X-ray luminosities by adopting a
standard neutron star cooling scenario, and by treating self-consistent
polar cap heating in the Harding \& Muslimov (1998) model, respectively.
Our results show that the observed different dependences of the high
energy luminosities on the pulsar spin-down luminosities, i.e., 
$L_\gamma \propto (L_{\rm sd})^{1/2}$ and $L_x \sim 10^{-3} L_{\rm sd}$, 
are well reproduced. We found that, for normal pulsars, both the hard 
(ASCA band) and the soft (ROSAT band) X-ray luminosities are dominated 
by the non-thermal X-rays of ICS origin, although for some pulsars, 
thermal components due to either neutron star cooling or polar cap 
heating can have comparable luminosities so that they are detectable.
For the millisecond pulsars, our predicted upper limits of
the thermal luminosities due to polar cap heating are usually higher 
than the ICS-origin non-thermal components if there are no strong
multipolar magnetic field components near the neutron star surface, 
thus the {\em pulsed} soft X-rays in the ROSAT band from most of the 
millisecond pulsars might be of thermal origin.

\end{abstract}  
  
  
\section{Introduction}  
Among about 1000 spin-powered radio pulsars detected so far, 35 are
detected as X-ray sources with 11 of them being X-ray pulsars
(for recent reviews, see Becker \& Tr\"umper 1997, update version
1999, hereafter BT97; Saito 1998, hereafter S98; Saito et al. 1997),
and at least 8 are detected also as $\gamma$-ray pulsars (for recent
reviews, see Thompson et al. 1997; Thompson et al. 1999. The eighth
one, PSR 1046-58, was recently reported by Kaspi et al. (1999)). Though
detailed data for these spin-powered high energy pulsars are still
poor, and some of them even have a great diversity of emission
features, there do exist some empirical laws. One prominent
feature is the dependence of the observed high energy luminosities
on the pulsar spin-down luminosities: For $\gamma$-rays,
it is found that $L_\gamma\propto (L_{\rm sd})^{1/2}\propto B/P^2$
(Thompson et al. 1997); while for X-rays, the dependence of $L_{x}
(ROSAT)\sim 10^{-3} L_{\rm sd}\propto B^2/P^4$ is found in ROSAT
(0.1-2.4 keV) data (BT97). A rough dependence of $L_{x}(ASCA)\propto
(L_{\rm sd})^{3/2}\propto B^3/P^6$ for ASCA (0.7-10 keV) data is
reported by Saito (1998, hereafter S98) and Saito et al. (1997) based
on a relatively smaller sample than the ROSAT data. However, in view
of the large uncertainties such as unknown beaming, hydrogen column
density, distance, assumptions of spectral shape, instrument response,
calibration uncertainties, and so on, the ASCA data are not inconsistent
to the ROSAT data.

Combining the ROSAT and ASCA data, spectral analysis of some of the
X-ray pulsars are available. Full-surface thermal emission
components are identified from four pulsars, i.e., Vela (\"Ogelman, 
Finley \& Zimmermann 1993), Geminga (Halpern \& Wang 1997), PSR 0656
+14 and PSR 1055-52 (Greiveldinger et al. 1996; Wang et al. 1998),
usually accompanied by a hard non-thermal component. 
As for the small-area hot thermal 
components, the evidence is not robust. Geminga (Halpern \& Ruderman 
1993), PSR 0656+14 and PSR 1055-52 (Greiveldinger et al. 1996) have
been reported to show such components, but follow-up data analyses
show that the hard components are more likely power-law (Halpern
\& Wang 1997; Wang et al. 1998). Two nearby old pulsars, PSR 1929+10
and PSR 0950+19 are reported to be dominated by a single small-area
hard thermal component (Wang \& Halpern 1997), but such conclusions
are also not solid, since PSR 1929+10 could be equally well fitted by 
a single power law (BT97), and the measurements of PSR 0950+19 reported
by Wang \& Halpern (1997) are heavily contaminated by another source
in the ASCA field of view (S98). For the millisecond pulsar PSR
J0437-4715, two-component spectral fits using the ROSAT plus EUVE
data require a hot small-area thermal component (Becker \&
Tr\"umper 1993; Halpern, Martin \& Marshall 1996; Zavlin \& Pavlov
1998), although a simple power-law (BT97) or a broken power-law
(Becker, Tr\"umper 1999) fit is also acceptable.
Among others, 7 ROSAT (BT97) and 1 ASCA (S98)
pulsars are identified to show power-law spectral features, which
suggests a non-thermal origin. The detected X-ray photons from the
rest of pulsars are too few for spectral fits, but they are commonly
thought to be of non-thermal origin as well, due to the close
relationships of their X-ray luminosities to the spin-down
luminosities (e.g. BT97; S98).
 
Two general classes of models have been proposed for high-energy
pulsars.  The polar cap models (Harding 1981; Daugherty \& Harding
1982; 1994; 1996, hereafter DH96; Sturner \& Dermer 1994; Sturner,
Dermer \& Michel 1995, hereafter SDM95) interpret pulsar high energy
emission as curvature radiation (CR) or magnetic inverse Compton
scattering (ICS) induced pair production cascade in the polar cap
region; while the outer gap models (e.g. Cheng, Ho \& Ruderman 1986a,b;
Romani \& Yadigaroglu 1995; Romani 1996; Zhang \& Cheng 1997)
interpret pulsar high
energy emission as fan beam radiation originating in the large vacuum
gaps in the outer magnetosphere beyond the null charge surface. In the
early polar cap cascade pictures, the observed $\gamma$-emission is
interpreted as the primary CR and the synchrotron radiation (SR) from
the higher generation pairs (Daugherty \& Harding 1982, 1994, 1996; 
Zhao et al. 1989; Lu \& Shi 1990;
Lu, Wei \& Song 1994 hereafter LWS94; Wei, Song \& Lu  
1997, hereafter WSL97). Recently, the importance of the ICS with the
soft thermal photons from the NS surface by the primary particles is
noticed. It is found that ICS is not only a significant energy loss
(Kardash\"ev et al. 1984; Xia et al. 1985; Daugherty \& Harding 1989;
Sturner 1995) mechanism, but is also a mechanism of $\gamma$-ray
emission (Sturner \& Dermer 1994; SDM95) and to ignite pair-production
cascades (Zhang \& Qiao 1996; Zhang et al. 1997a,b). However, Harding
\& Muslimov (1998, hereafter HM98) argued that the asymmetry of upward
versus downward ICS induced pair cascades make a ICS-controlled
accelerator usually unstable near the stellar surface in the
space-charge-limited flow acceleration model, so that a CR-controlled
accelerator could be stablized at a height of 0.5-1 stellar radii
above the NS surface when the energy loss efficiency of ICS is less
than that of CR. As a result, the CR-induced cascades picture (DH96)
could still be maintained, although ICS of the primaries will also
contribute to the final $\gamma$-ray spectrum. The high-altitude
accelerator naturally matches the extended polar cap cascade model
(DH96).

In such a polar cap model, it is predicted that there is a low-energy
turnover at the blueshifted local cyclotron frequency in the high
energy spectra of pulsars (Harding \& Daugherty 1999). Such a feature
is also noticed by Rudak \& Dyks (1999) who argued that the
non-thermal X-rays observed by ROSAT and ASCA pose a challenge to such
a model, and it could be a discriminator between the rivaling classes
of models, since the dependence of $L_x\sim 10^{-3}L_{\rm sd}$ could be
interpreted in terms of a thick outer gap model (Cheng, Gil \& Zhang
1998, hereafter CGZ98; Cheng \& Zhang 1999, hereafter CZ99). However,
there is yet another important point which has been overlooked in
previous studies of polar cap cascade scenarios. This is the possible
contributions of the ICS photons produced by higher generation pairs
to the pulsar high energy radiation.  Even in the ICS-induced polar
cap cascade model (SDM95), these contributions were not included,
since the authors found that these ICS photons are not energetic
enough to pair-produce again
\footnote{This is true for most of pulsars, but is not the case for
the pulsars with high magnetic fields. See our calculation results
presented in Table 2 and corresponding discussions.},
though they admitted that most of these
scatterings occur in the ``resonant'' regime so that they should be
important for spectral formation. Here we will improve the polar cap
cascade picture by including the ICS contributions of the higher
generation pairs. We will adopt an analytic approach to describe
such a more complicated ``full-cascade'' process and examine how
significant these contributions are. A straightforward conclusion is
that, the low energy tail of such ICS spectra can naturally provide a
non-thermal X-ray component, which could contribute to the ROSAT and
ASCA bands observations. We will show that, the full cascade model
combined with the HM98 acceleration model can not only present a 
clear $L_\gamma \propto (L_{\rm sd})^{1/2}$ feature, but the ICS-origin
non-thermal X-ray components with the combination of the thermal 
components due to polar cap heating can also reproduce the rough
$L_x \sim 10^{-3} L_{\rm sd}$ dependence. The preliminary results
of this study is shown in Zhang \& Harding (1999).
 
The organization of this paper is as follows. In Sect. 2.1-2.4, we
describe in detail the analytic approach to delineate the cascade
processes. Such descriptions are independent of the
acceleration model and thus could be used in any model when the inner
accelerator (gap) features are specified. Sect. 2.5 reviews the main
features of the acceleration model proposed by Harding \& Muslimov
(HM98), which has incorporated the frame-dragging-induced parallel
electric fields and both the CR- and ICS- induced cascade processes,
and therefore is hitherto the most complete space-charge-limited flow
acceleration model. In Sect. 3.1-3.4, we explicitly present our model
predictions of the total high energy luminosities (mainly
the $\gamma$-ray luminosites), thermal, and non-thermal
X-ray luminosities from the known pulsars, and compare them with the
observations from CGRO, ROSAT and ASCA. Our main results are
summarized in Sect.4 with some discussions.

\clearpage 
\section{The Model} 
 
\subsection{Resonant scattering condition and the ``full-cascade'' 
 picture \label{full}}
 
In our work, it is assumed that there exist two thermal emission
components from neutron stars: one soft component over the whole
surface with the temperature $T_s$, and a hot component at the
polar cap region with the temperature $T_h$. As reviewed in
the previous section, these components have actually been
detected from some pulsars (though with some uncertainties), 
and they are commonly expected by various theories. The
full surface radiation is predicted in various neutron star cooling
(e.g. Van Riper \& Lamb 1981; Nomoto \& Tsuruta 1987) and internal
heating (e.g. Alpar et al. 1984; Shibazaki \& Lamb 1989; Schaab et
al. 1999) models. In the outer gap heating models, it is supposed that
the hard X-rays emitted from the hot polar cap may be reflected back
to the surface by a ``pair-blanket'' due to the magnetic cyclotron
resonant absorption, so that they might be reemitted from the full
neutron star surface with softer energies (Halpern \& Ruderman 1993;
Zhang \& Cheng 1997; Wang et al. 1998; CGZ98; CZ99). A hotter polar
cap is expected by either polar cap heating (e.g. Arons 1981;
Harding, Ozernoy \& Usov 1993; Muslimov \& Tsygan 1992; HM98) or 
outer gap heating (again Halpern \& Ruderman 1993, etc.).  The 
ICS of the relativistic particles with these thermal
photons is the main topic discussed in this paper. Thus our present
model is solely contingent upon the existence of the thermal photon 
fields in the neutron star vicinity, and can not be justified if it
turns out that there is no such thermal emission at all. Note there
should be also some ICS processes with the soft non-thermal X-rays
if they do exist, and such processes will not be discussed in this
paper.

The so called ``resonant scattering condition''
(Daugherty \& Harding 1989; Dermer 1990) is essential in our
``full-cascade'' picture. This condition ensures that the bulk of the
soft thermal photons near the neutron star surface with peak energy
$2.82 kT=0.24 T_6$(keV) could be Lorentz-boosted to the particle's
rest-frame local field cyclotron energy $\hbar\omega_{_B}=\hbar 
eB_e/mc= 11.6 B_{e,12}$(keV), where $B_e$ is the magnetic field
strength at the emission (scattering) point. More specifically, 
this condition reads $\gamma (1-\beta\mu) \cdot 2.82 kT
\gtrsim \hbar \omega_{_B}$, or  
\begin{equation} 
\gamma\gtrsim\gamma_{\rm res}={48 B_{12} \over [(1-\beta\mu)T_6]},
\label{res1} 
\end{equation} 
where $\mu=\cos\theta$ is the cosine of the incident angle of the
scattering, and $\gamma$ is the Lorentz factor of the electron.
For the case of actual pulsars with two thermal
components, we adopt different geometries with
\begin{equation} 
\mu_{\rm s,max}=(1-r_{e,6}^{-2})^{1/2} 
\end{equation} 
for the soft full-surface component, and
\begin{equation} 
\mu_{\rm h,max}={\rm Max}\left[{r_{e,6}-1 \over \left[
(r_{e,6}-1)^2+\left({\Omega R \over c}\right)^{1/2}\right]
^{1/2}}, (1-r_{e,6}^{-2})^{1/2}\right]
\label{muh} 
\end{equation} 
for the hot polar cap component, where $r_{e,6}$ is the radius
of the emission point in units of $10^6$cm (i.e. the stellar radius). 
Note that the two terms on the right hand of (\ref{muh}) are for
different emission height regimes, the conjuction of which is the
height above which the horizon of the emission point is greater 
than the polar cap area. 
Then the resonant condition (\ref{res1}) becomes
\begin{equation} 
\gamma\gtrsim\gamma_{\rm res}={48 B_{12}\over {\rm Max}[(1-\beta\mu_
{\rm s,max})T_{s,6}, (1-\beta\mu_{\rm h,max})T_{h,6}]}.
\label{res} 
\end{equation} 
 
In the resonant ICS regime, the e-folding mean free path of the
particles with Lorentz factor $\gamma$ (Dermer 1990; Sturner 1995)
\begin{equation} 
\begin{array}{ll} 
\lambda_{\rm res}& \simeq -0.061 ({\rm cm})\gamma^2 T_6^{-1}
B_{e,12}^{-2} \\ 
& \times \ln^{-1}\left[1-\exp\left(-{134 B_{e,12}\over \gamma T_6
(1-\mu_c)}\right)
\right], 
\end{array} 
\label{lambdares} 
\end{equation} 
is very small due to the tremendous enhancement of the scattering
cross section near the resonant frequency, so that the particles could
efficiently convert their kinetic energies into radiation before
moving a long distance. Thus, as long as the condition (\ref{res}) is
satisfied, the Lorentz factor of the particles will keep decreasing
all the way down to the value when the condition (\ref{res})
fails. Then, most of the kinetic energies of the particles are
converted to radiation.
 
According to Dermer (1990) and Sturner (1995), there is a wide regime
of the Lorentz factors of the particles which satisfies the resonant
condition (\ref{res}). The resonant scattering takes place not only
for the primary particles within the accelerator, but also for the
secondary pairs produced above the accelerator. Zhang et al. (1997b)
have calculated the Lorentz factor evolution above the gap for the
secondary pairs within the framework of the Ruderman-Sutherland (1975,
hereafter RS75) type vacuum gap acceleration model. They found that
the resonant ICS process can reduce the energies of the secondary
pairs to much lower values than the original ones, which may constrain
pulsar radio emission frequency to below a certain value. 
However, they did not consider the
contribution of these energies to the high energy radiation of the
pulsars. In fact, such secondary ICS processes will also produce high
energy emission which contributes to the final spectra of the
$\gamma$-ray pulsars. The low frequency tail of such spectra can
contribute to a non-thermal X-ray emission component. As a result, it
is important to modify the conventional CR(or ICS)-SR polar
cap model to include this ICS effect. 
 
We then come to a ``full-cascade picture'' (Fig.1): the primary
particles within the accelerator emit primary $\gamma$-rays, which
produce secondary pairs via $\gamma-B$ process. The
perpendicular energies of these secondary pairs are rapidly converted
to radiation via SR, while most of the parallel energies will also be
converted to radiation via ICS if the resonant scattering condition
(\ref{res}) is satisfied. Thus two-branch secondary generation
$\gamma$-rays are produced. These secondary $\gamma$-rays
may undergo the $\gamma-B$ process again to
produce higher two-branch generation pairs and $\gamma$-rays. Such
processes go on and on until the $\gamma$-rays escape from the
magnetosphere.
 
\centerline{}
\centerline{\psfig{file=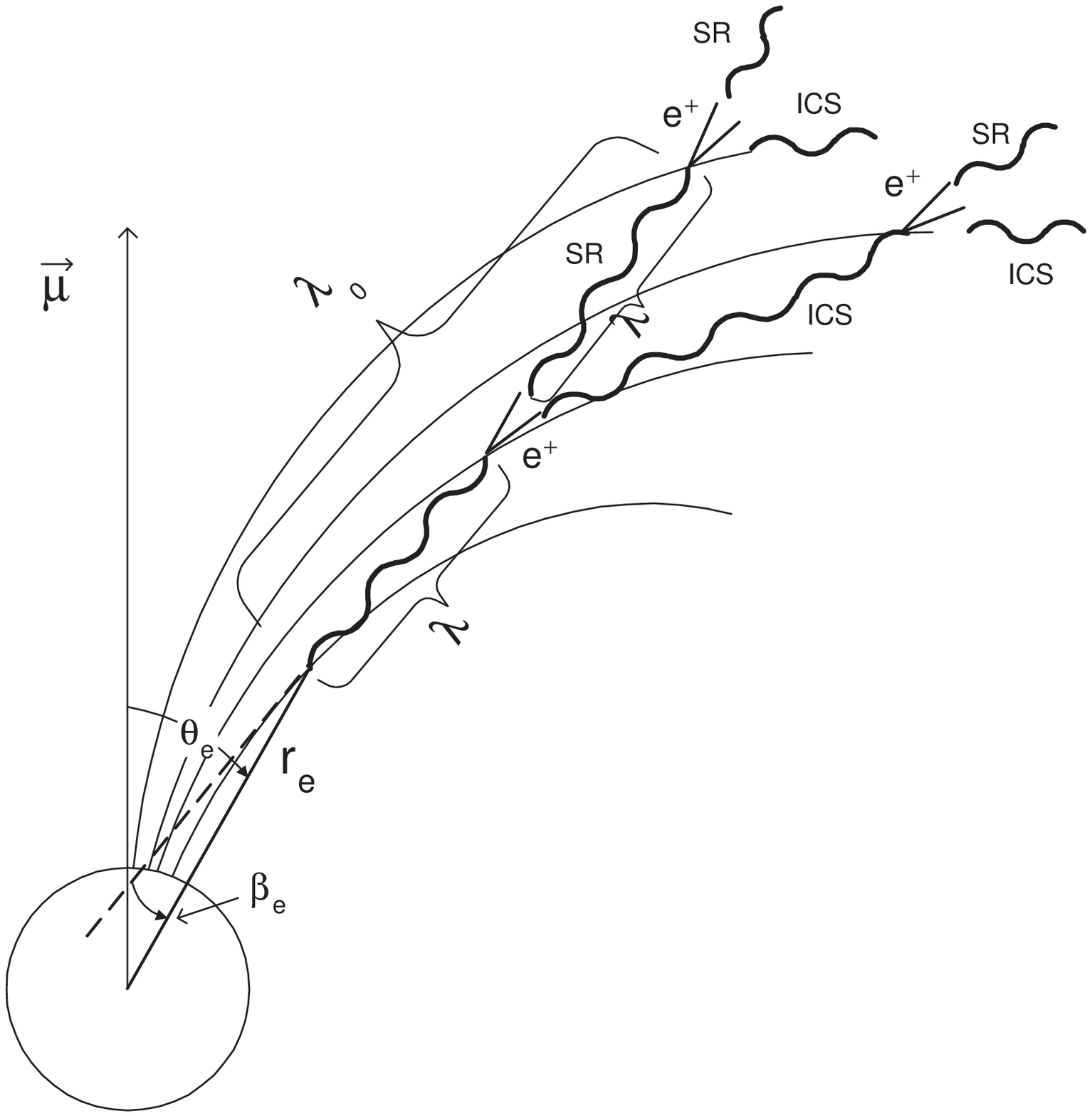,width=9.0cm}} 
\vspace {2.0cm}
\figcaption{A Schematic illustration of the full cascade process. The  
coordinates of the emission point are ($r_e$,$\theta_e$), and $\beta_e$  
is the angle between the radial and tangential directions at the emission 
point. The wavy lines denote the photons. SR-branch photons follow the 
original direction of the pairs, while ICS-branch photons follow the 
tangential direction. $l$ is the attenuation length of each generation 
photon, and $l_0$ is scaled from the original emission point where the 
photon is emitted tangentially.
\label{cartoon}  }
\centerline{}
 
In this paper, we will limit ourselves to an analytic approach to
understand the complex ``full cascade'' processes. An analytic method
(e.g. LWS94; WSL97) is helpful to present an over-all picture for the
pulsar population, and can, to some extent, predict some features of
the numerical results.
 
We have some remarks on the validity of condition (\ref{res}) for
millisecond pulsars.  Since the cyclotron energies in millisecond
pulsars are much lower than the thermal peak energies, very small
incident angles (e.g. $\mu \sim 1$) are required to keep the thermal
photons Lorentz-down-boosted to the cyclotron frequency, and we can
get an absolute upper limit of Lorentz factors of the particles for
resonant scattering by adopting $\mu=1$. Noticing $1-\beta \simeq
1/2\gamma^2$, this turns out
\begin{equation} 
\gamma \leq 105 T_6/B_8. 
\label{msres} 
\end{equation} 
When this condition fails, as long as condition (\ref{res}) is
satisfied, there are still some off-peak thermal photons which are
resonantly scattered, but the energy conversion efficiency is much
lower due to the smaller ``soft'' photon numbers.  Even if both
(\ref{res}) and (\ref{msres}) are satisfied, the mean free path of
resonant scattering (\ref{lambdares}) is still much longer than the
normal pulsars due to the tremendous decrease of the field
strength. Thus ICS energy conversion efficiency is low for millisecond
pulsars. As shown later in Sect.\ref{nonth-x}, for the millisecond 
pulsars, usually the non-thermal soft X-rays of ICS-origin are 
dominated by the polar-cap-heating-produced thermal X-rays.
 
\subsection{$\gamma-B$ pair production and photon escaping energy} 
 
The one-photon pair production process or the $\gamma-B$ process plays
an essential role in the polar cap cascade scenario. A detailed
description of the process and explicit numerical computations were
presented in Harding, Baring \& Gonthier (1997, hereafter HBG97). Here
we will adopt some approximations within certain reasonable regimes and
come to some analytic expressions.
 
\subsubsection{Attenuation mean free path} 
We adopt the simple asymptotic formula of Erber (1966). Assuming that
the one photon attenuation coefficient is constant over the
distance interested in, we get the mean free path of attenuation,
which is simply the inverse of the attenuation coefficient (e.g. RS75)
\[ 
l={4.4\over (e^2/\hbar c)}{\hbar\over mc}{B_c\over B_\perp}\exp
\left({4\over 3\chi}\right)
\] 
or
\begin{equation} 
l_6=1.03B_\perp^{-1}\exp\left({4\over 3\chi}\right),
\label{l6} 
\end{equation} 
where
\begin{equation} 
\chi={1\over 2}{E_\gamma \over mc^2}{B_\perp\over B_c}={1\over 2}
\epsilon_\gamma B'_\perp ,
\label{chi} 
\end{equation} 
and $B_c=m^2c^3/e\hbar=4.414\times 10^{13}$ is the critical magnetic
field, $B'_\perp=B_\perp/B_c$, $B_\perp =B \sin\theta_{\rm kB}$,
$E_\gamma$ is the energy of the $\gamma$-photon in the local rest
inertia frame, which should be corrected by a redshift factor (see
eq.[\ref{redshift}]) when compared with the observations,
$\epsilon_\gamma$ is $E_\gamma$ normalized by $mc^2$, and $\theta_{\rm
kB}$ is the angle between the photon and the field directions. Such an
asymptotic formula is accurate enough when $B'=B/B_c\leq 0.1$ (Daugherty 
\& Harding 1983), which is the case for most pulsars in the HM98 model
since the accelerators have been lifted to higher altitudes where
strength of the local magnetic fields have dropped nearly an order of
magnitude with respect to the surface fields (see Sect.\ref{HM98}). But 
we should bear in mind that the asymptotic formula is unreliable at near
threshold regime ($\epsilon_\gamma \sin\theta_{\rm kB} \gtrsim 2$),
and should be modified to a more accurate asymptotic form (Daugherty
\& Harding 1983)
\begin{equation} 
l_6=1.03B_\perp^{-1}\exp\left[{4\over 3\chi}(1+\delta)\right]
\end{equation} 
with $\delta=0.42(\epsilon_\gamma/2 \sin\theta_{\rm kB})^{-2.7}
B'^{-0.0038}$, or even no asymptotic formula at all (HBG97).
 
\subsubsection{Geometric relations} 
The importance of general relativistic effects on one-photon pair 
production attenuation in a Schwarzschild metric has been examined by 
Gonthier \& Harding (1994, hereafter GH94) and HBG97. They found that 
curved spacetime effects may influence the process in three ways, namely 
1) curvature of the photon trajectories, 2) the redshift of the photon
energy, and 3) the change in the dipole magnetic field (the shape of
the field configuration and the enhancement of the field
strength). Their conclusion is that the distortion of the dipole in
curved spacetime is compensated by the curved photon trajectory
effect, but the increasing of the field strength in curved space-time
will enhance the $\gamma-B$ absorption and hence, lower the escaping
photon energy. The relative redshifts of the photon energies between
different generations are not large since the height differences
between different generations are usually much smaller than one stellar
radius. However, the redshift effect should be taken into account when
comparing the calculated frequencies (in the local inertia frame) with
the observational frequencies (the observers are located at infinity)
(see eq.[\ref{redshift}]). In our analytic study below, we will keep
the flat spacetime geometry, i.e. assuming strict dipolar field
configuration and straight trajectories of photon propagation. Neither
of these assumptions are precise enough, but a combination of the two
comes to a good approximation (GH94). We only incorporate an
enhancement factor of the field strength due to curved spacetime (see
eq.[\ref{q}]). The redshift effect is only taken into account when
comparing theoretical results with the observations
(see eq.[\ref{redshift}]).
 
In an exact ``dipole'' field in flat space-time, suppose a
particle at $(r_e, \theta_e)$ emits a photon along the field line,
then there is a pure geometric relation between the perpendicular
field strength the photon encounters, i.e. $B_\perp$, and the travel
distance of the photon $l$ (Hardee 1977)
\begin{equation} 
\begin{array}{ll} 
B_\perp & = {3\over 4}B_p R^3 {\left[{l\sin\theta_e \over (r_e^2
+2r_el\cos\beta_e+l^2)^{5/2}}\right]} \\ & \times
[r_e(1+\sin^2\beta_e)+l\cos\beta_e],
\end{array} 
\label{Bperp1} 
\end{equation} 
where $\beta_e$ is the angle between the field direction and the radial
direction of the emission point, with $\tan \beta_e=1/2\tan\theta_e$
(see Fig.1). Note $B_p R^3 /2$ (rather
than $B_p R^3$) is the dipole magnetic moment of a uniformly
magnetized sphere of radius $R$ (Usov \& Melrose 1995; HBG97), so that
$B_p$ is identical to the surface magnetic field $B_s$, and is
connected with pulsar period $P$ and period derivative $\dot P$ by
\begin{equation} 
B_p=6.4 \times 10^{19} {\rm G} (P \dot P)^{1/2},
\end{equation} 
which is twice of the value adopted by Manchester \& Taylor (1977) and
Michel (1991).
 
Near the polar cap region, one has $\theta_e\ll 1$, and equation
(\ref{Bperp1}) becomes
\begin{equation} 
B_\perp\simeq {3\over 4}B_p R^3{l\theta_e\over r_{ a}^4}= {3\over 4}
B_{ a} \theta_e{l\over r_{ a}},
\label{Bperp2}  
\end{equation} 
where subscript ``a'' indicates the ``absorption point'', so that
$r_{a}\simeq r_e+l$ and $B_a=B_p\left({R\over r_a}\right)^3$. Such a
$B_\perp$ has a maximum when $l\simeq r_e/3$ (Hardee 1977), with
\begin{equation} 
B_{\rm \perp,max}\simeq 0.08B_p {\left({R\over r_e}\right)^3}\theta_e
=0.08B_e\theta_e
\label{Bmax1} 
\end{equation} 
or
\begin{equation} 
B_{\rm \perp,max}\simeq1.2\times 10^9{\rm G}B_{
e,12}P^{-1/2}r_{e,6}^{1/2}\xi.
\label{Bmax2} 
\end{equation} 
Here $B_p$, $B_e$, $B_{ a}$ are the magnetic fields at the surface,
the emission point, and the absorption point, respectively,
$\theta_e=\xi\left({\Omega r_e\over c}\right)^{1/2}$ 
(according to the dipolar configuration),  
and $0<\xi=\theta_s / \theta_{\rm pc} <1$ is the ratio of the field
line magnetic colatitude at the surface, $\theta_s$, to the polar cap
angle at surface, $\theta_{\rm pc}=\left({\Omega R\over
c}\right)^{1/2}$.
 
For $l \ll r_a$ (this is the case for most of the generations except
for the last one which will not pair produce again, see numerical
results of HBG97, thus is a plausible approximation to describe the
averaged cascade processes), we have $r_a\simeq r_e$, $B_a\simeq B_e$,
and equation (\ref{Bperp2}) reads
\begin{equation} 
B_\perp \simeq {3\over4}B_e {l \over r_e/\theta_e}=B_e {l\over
\rho_e},
\label{Bperp3} 
\end{equation} 
where
\begin{equation} 
\rho_e \simeq {4\over 3}{r_e \over \theta_e}=9.2\times 10^7({\rm cm})
r_{e,6}^{1/2} P^{1/2} \xi^{-1}
\label{rho} 
\end{equation} 
is the curvature radius of field line at the emission point
(assuming dipolar configuration). Equation
(\ref{Bperp3}) is identical with the approximation used in many
previous analytic studies (e.g. RS75).
 
We now incorporate the correction of the field enhancement effect in
curved spacetime. By taking into account the general relativistic
effect, the magnetic field measured in a local rest frame of reference
in curved spacetime reads (see Wasserman \& Shapiro 1983 and GH94, but
adopting magnetic moment $B_p R^3/2$, however.)
\begin{equation} 
\begin{array}{ll} 
{\bf B}_{\rm curved} & =B_{\rm r,curve} \hat r +B_{\rm
\theta,curve}\hat\theta \\ & = - {3 B_p R^3 \cos\theta\over r_g^2 r}
\left[{r\over r_g}
\ln \left(1-{r_g\over r}\right)+1+{1\over2}{r_g\over r}\right] \hat r\\ 
& + {3 B_p R^3\sin\theta\over r_g^2 r}\left[\left( {r\over r_g}-1
\right) \ln
\left( 1-{r_g\over r} \right) +1-{1\over 2}{r_g\over r} \right]  \\ 
& \times \left(1-{r_g\over r}\right)^{1/2} \hat \theta,
\end{array} 
\end{equation} 
where $r_g={2GM \over c^2}$ is the Schwarzschild radius. Let us
overlook the modification of the ``dipole'' configuration of this
formula (the effect is actually almost canceled by the effect of the
curved photon trajectory, GH94), and compare it with the magnetic
field expression in the flat spacetime (note also adopting $B_p R^3/2$
as the magnetic torque)
\begin{equation} 
\begin{array}{ll} 
{\bf B}_{\rm flat} & =B_{\rm r,flat}\hat r +B_{\rm
\theta,flat}\hat\theta \\ & = {B_p R^3\cos\theta \over r^3} \hat r +
{B_p\sin\theta \over 2r^3}\hat\theta
\end{array} 
\end{equation} 
We see the curved spacetime effect will increase the strength of the
magnetic field by a factor of
\begin{equation} 
q(r)={[B_{\rm r,curved}^2(r)+B_{\rm \theta,curved}^2(r)]^{1/2} \over
[B_{\rm r,flat}^2(r)+B_{\rm \theta,flat}^2(r)]^{1/2}}.
\label{q} 
\end{equation} 
With this correction, equations
(\ref{Bperp1},\ref{Bperp2},\ref{Bmax1},\ref{Bmax2}) should be modified
by multiplying a factor of $q(r)$. Note that the factor $q(r)$ is
equivalent to the function $f(\eta)$ introduced in HM98, which 
decreases with $r$, and converges to unity at infinity.
For a typical neutron star with $M=1.4M_\odot$ and $R=10$km,
we have $r_g/R\simeq 0.4$, thus the modification factor at the surface
is $q(R)\sim 1.4$, and $q(r)\sim (1.18-1.25)$ for the height of
0.5-1 stellar radii, which is the interested region discussed in
this paper (see details in Sect.\ref{HM98}).  
Then the maximum perpendicular magnetic field (eq.[\ref{Bmax2}]) now
should be modified by multiplying a factor $q(r_e)$, which is
\begin{equation} 
B_{\rm \perp,max,curved}({\rm nPSR}) \simeq 1.4\times 10^9{\rm G}B_{
e,12}P^{-1/2}r_{e,6} ^{1/2}\xi
\label{Bmax3} 
\end{equation} 
for normal pulsars, and
\begin{equation} 
B_{\rm \perp,max,curved}({\rm msPSR}) \simeq 5.1\times 10^7{\rm G}B_{
e,9}P_{-3}^{-1/2}r_{e,6} ^{1/2}\xi
\label{Bmax3'} 
\end{equation}
for millisecond pulsars. 
Here $B_{e}=B_{p}\left({R\over r_e}\right)^3$ is still the
field strength at the emission point in {\em flat} spacetime.
 
There is yet another effect, namely the ``non-zero pitch angle''
effect, that will reduce the attenuation length of the SR photons,
which has been overlooked in some of the previous studies.
The expressions
of $B_\perp$ (eqs.[\ref{Bperp1},\ref{Bperp2},\ref{Bperp3}]) are
derived assuming that photons are emitted along the tangent direction
of the field line, which is not the case for higher generation SR
photons which follow the direction of the pairs which have a non-zero
pitch angle with respect to the field lines. Thus
equations (\ref{Bperp1},\ref{Bperp2},\ref{Bperp3}) have in fact
underestimated the strength of the perpendicular field. However, the
geometric relation in (\ref{Bperp1}) could still be used by replacing
the emission point ($r_e$, $\theta_e$) with the point ($r_{e,0}$,
$\theta_{e,0}$) where the $\gamma$-rays are emitted tangentially. Defining
$l_0$ as the distance between ($r_{e,0}$, $\theta_{e,0}$) and ($r_a$,
$\theta_a$), and $l$ as the mean free path of the SR
$\gamma$-rays, i.e. the distance between ($r_e$, $\theta_e$) and
($r_a$, $\theta_a$) (see Fig.1 for illustration), again adopting
$\theta_e \ll 1$, the perpendicular field is then
\begin{equation} 
B_\perp = B_{ a}\theta_e{3r_e^2 l_0 \over 2r_{ a}
(r_e+l_0-l)(2r_e+l_0-l)},
\label{Bperp4} 
\end{equation} 
which is simply
\begin{equation} 
B_\perp \simeq B_e {l_0 \over \rho_e}
\label{Bperp5} 
\end{equation} 
when $l<l_0 \ll r_e$. This is just equation (\ref{Bperp3}) by changing
$l$ to $l_0$.  We see the higher the SR generation, the stronger the
perpendicular fields, since $l_0$ increases with generations. This
effect will raise the value of $\chi$ for pair-production with the
same $l_6$ (see Sect.\ref{chi?} and Fig.2) and hence, tend to 
increase the number of the SR generations.
Such a ``non-zero pitch angle'' effect is non-relevant to the ICS
branch, however, since the ICS photons are emitted by the pairs after
emitting their perpendicular energies via SR, and hence, follow the
tangential direction of the field line at the emission point. 
 
\subsubsection{How large is $\chi$ for pair production? \label{chi?}} 
Equation (\ref{l6}) can give a self-consistent $l_6$ and $\chi$ for
one-photon pair production. It is essential that the parameter $\chi$
does not sensitively depend upon various parameters, since small
changes in $\chi$ correspond to exponentially large changes of
$l_6$. Such a feature makes some simple analytic formulae available in
analytic studies (e.g. RS75; LWS94; WSL97). In these models, $\chi$ is
adopted as 1/15. However, in order to treat the process more
precisely, it is still worth examining the possible variation of
$\chi$ for different pulsar parameters. A smaller $l_6$ can increase
$\chi$ and hence increase the total number of generations (i.e. the
generation order parameter, LWS94, WSL97, also see Sect.\ref{gop} and
\ref{zeta}). In fact, Zhang et al. (1997b) has adopted 
$\chi=0.09, 0.12$ to estimate the parameters of different 
polar gap modes.

Using (\ref{l6}), (\ref{rho}) and (\ref{Bperp3}), one can get a more
accurate value of $\chi$ when $B_e$, $P$, $r_{e,6}$ (these three
parameters are specified for a certain pulsar and an acceleration
model), $l_6$ (this parameter depends on the $\gamma$-ray energy) and
$\xi$ (depends on the field line) are given
\begin{equation} 
\chi={4\over 3}\ln^{-1} [0.01 B_e P^{-1/2} l_6^2 r_{e,6}^{-1/2} \xi] 
\label{chi1} 
\end{equation} 
Figure 2 plots the dependence of $1/\chi$ on $l_6$ (corresponding to
different $E_\gamma$) for two sets of parameters: one set is for a
typical normal pulsar with $B_e = 10^{12}$G, $P=0.1$s, $r_{e,6}=1.8$
and $\xi=1$; another is for a typical millisecond pulsar with
$B_e=5\times10^8$G, $P=0.005$s, $r_{e,6}=1.2$ and $\xi=1$. Note that
different adoptions of $r_{e,6}$ are based on the HM98 model (see
Sect.\ref{HM98}). We found that, the parameter $\chi$ has
deviated from 1/15 for most of the cases. For most of the generations,
we have $l_6 \ll 1$ (see numerical results in HBG97), thus $\chi$ is
larger than 1/15.  For millisecond pulsars, the values of $\chi$ are
even larger.

\centerline{}
\centerline{\psfig{figure=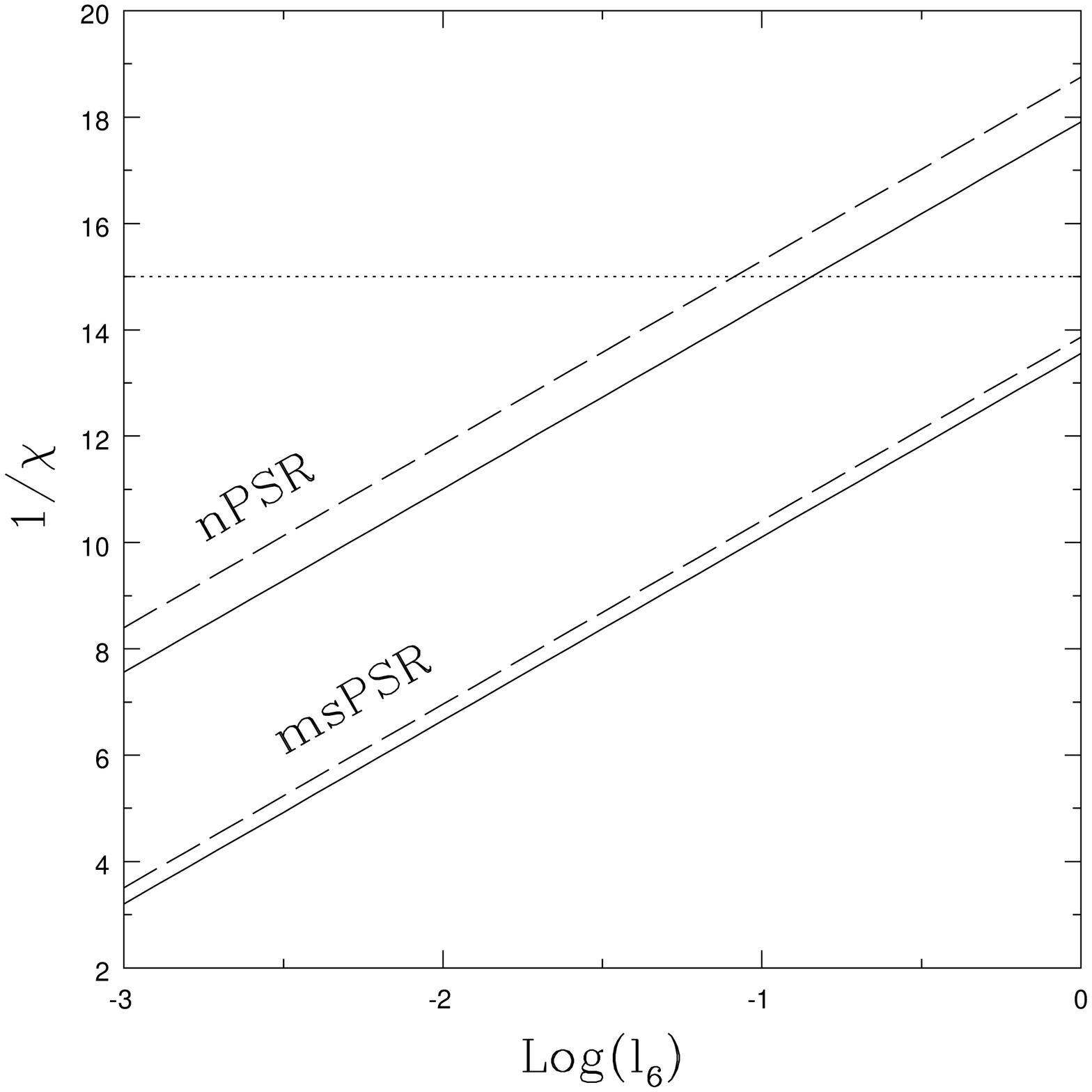,width=8.0cm}}
\figcaption{The $l_6$ dependence of the critical value of $\chi$ for 
$\gamma-B$ pair production ($1/\chi$ is plotted). The upper two lines 
are for a typical normal pulsar with $P=0.1$s, $B_{e,12}=1.0$, $r_{e,6} 
=1.8$ (thus $B_{p,12}=5.8$), and $\xi=1$; the lower two lines are for a  
typical millisecond pulsar with $P=0.005$s, $B_{e,12}=5.0\times 10^{-4}$, 
$r_{e,6}=1.2$ and $\xi=1$ (thus $\log B_p=8.9$). The dashed lines are 
for the case of `tangentially' emitting, and the solid lines are for 
the case with an estimate correction of the non-zero pitch angle effect. 
The dotted line marks the conventional value $\chi=1/15$. 
\label{fig:chi}}
\centerline{}
  
The critical $\chi$ for pair production could be further enhanced by
including ``non-zero pitch angle'' effect as discussed above. With
(\ref{Bperp5}) rather than (\ref{Bperp3}), the equation (\ref{chi1}) now
becomes
\begin{equation} 
\chi={4\over 3}\ln^{-1} [0.01 B_e P^{-1/2} l_6 l_{0,6} r_{e,6}^{-1/2} 
\xi] 
\label{chi2} 
\end{equation} 
At higher generations, $l_{0,6}$ is larger than $l_6$. We have adopted
$l_{0,6}=3 l_6$ for normal pulsars and $l_{0,6}=1.5 l_6$ for
millisecond pulsars to cover this effect. The different adoptions
reflect different generation numbers in normal and millisecond
pulsars. The results are also shown in Fig.2.
 
With the results in Fig.2, we come to some ``recommended'' values for
the critical $\chi$. When calculating photon escaping energy (see
Sect.\ref{Eesc}), it is appropriate to adopt a smaller $\chi$ since
$l_6$ of the last generation $\gamma$-rays is usually around unity
(recall $B_{\rm max}$ is achieved when $l\simeq r_e/3$, also see
numerical results of HBG97). However, when describing the cascade
process, a larger $\chi$ should be adopted, since the averaged $l_6$
is much less than unity. In our following analytic formulae, we'll
adopt
\begin{equation} 
\chi_{\rm esc}{\rm (nPSR)} \sim 1/16, ~~~ \chi_{\rm cas}{\rm (nPSR)} 
\sim 1/12 
\label{chipsr} 
\end{equation} 
for normal pulsars, and
\begin{equation} 
\chi_{\rm esc}{\rm (msPSR)} \sim 1/12,  ~~~ \chi_{\rm cas}{\rm (msPSR)} 
\sim 1/8 
\label{chimspsr} 
\end{equation} 
for millisecond pulsars, according to the results of Fig.2.
 
\subsubsection{Photon escaping energy \label{Eesc}} 
Photon escaping energy is the energy below which the magnetosphere is
transparent.  A numerical result of the escaping energy of photons is
shown in Fig.2 of HBG97.  With (\ref{chi}), adopting $B_\perp$ as
the $B_{\rm \perp,max}$ described in (\ref{Bmax3}) and (\ref{Bmax3'})
in which the correction of curved spacetime effect has been included,
also choosing the recommended $\chi_{\rm esc}$ in equations
(\ref{chipsr},\ref{chimspsr}), we get
\begin{equation} 
E_{\rm \gamma,esc}{\rm (nPSR)}\simeq 2.0{\rm GeV}B_{
e,12}^{-1}P^{1/2}r_{e,6}^{-1/2}
\xi^{-1}\chi_{_{1/16}} 
\label{esc1} 
\end{equation} 
for normal pulsars and
\begin{equation} 
E_{\rm \gamma,esc}{\rm (msPSR)}\simeq 73{\rm GeV}B_{
e,9}^{-1}P_{-3}^{1/2}r_{e,6}^{-1/2}
\xi^{-1}\chi_{_{1/12}} 
\label{esc2} 
\end{equation} 
for millisecond pulsars.  Note these relations are quite different
from equation (36) of Hardee (1977), who has adopted the incorrect
form of $\gamma$-ray absorption formula. As pointed out by Zheng,
Zhang \& Qiao (1998),  although the general formula of $\gamma$-ray
attenuation including both the perpendicular magnetic fields and
the electric fields derived by Daugherty \& Lerche (1975) is correct,
when applying it to the situation of pulsars, the contribution of
the induced electric fields are actually of no importance, since in
the static frame (rather than the co-rotating frame), the emission
direction is not strictly along the direction of the field line due
to Lorentz transformation. Hardee (1977) has used the Daugherty \&
Lerche's formula by assuming that the photons are emitted tangently
along the field line in the static frame, and thus gave a incorrect
larger absorption coefficient\footnote{Note that in Zheng, Zhang \&
Qiao (1998), there are some mis-comments on the work of Harding,
Tademaru, \& Esposito (1978), Harding (1981), and Daugherty \&
Harding (1982), who have correctly treated the absorption problem
by including the ``aberration of the emission directions'' in the
static frame.}. WSL97 obtained a similar equation (their eq.[10])
as our (\ref{esc1}), but they did not include the general
relativistic correction and the non-zero pitch angle effect.

\subsection{Recursion relations between different generations \label{gop}} 
 
There exist some simple recursion relations between the characteristic
emission frequency of both the two-branch filial-generation (SR and
ICS) photons and the parent-generation photons. Consider a certain
parent generation $\gamma$-ray with dimensionless energy $\epsilon_{
i}$ (the origin of this photon is not important), for SR branch, the
typical energy of the filial-generation $\gamma$-rays is
\begin{equation} 
\epsilon_{i+1,{\rm SR}}={3\over 2}\gamma_{ i+1}^2 B'_e \sin\theta_{\rm kB} 
={3\over 4}\chi \epsilon_{ i}
\label{SR} 
\end{equation} 
where $\gamma_{i+1}=\epsilon_i/2$ is the Lorentz factor of the
filial-generation pairs following the direction of the parent photon,
which impacts the field with an angle of $\theta_{\rm kB}$. Define
$\kappa=\epsilon_{i+1}/\epsilon_i$ as the energy reduction factor of
the adjacent generations, we have (LWS94; WSL97)
\begin{equation} 
\kappa_{\rm SR}={\epsilon_{i+1,{\rm SR}} \over \epsilon_i}= 
{3\over 4}\chi. 
\end{equation} 
Recalling the recommended $\chi_{\rm cas}$ values in equations
(\ref{chipsr},\ref{chimspsr}), we have
\begin{equation} 
\kappa_{\rm SR}{\rm (nPSR)}\sim 1/16 
\label{kappasr1} 
\end{equation} 
for normal pulsars and
\begin{equation} 
\kappa_{\rm SR}{\rm (msPSR)}\sim 1/11 
\label{kappasr2} 
\end{equation} 
for millisecond pulsars. Note in (\ref{SR}) we have adopted the
classical formula to describe the SR process. In the strong magnetic
fields, however, the SR process should be strictly described
quantum-mechanically as the transitions between Landau levels with adjacent
energy interval of $\Delta E_L\simeq\hbar\omega_{_B}=11.2 B_{e,12}$keV
(for a review, see Harding 1991). The comparisons of the quantum and
classical treatments of the SR process have been presented in Harding
\& Preece (1987) and Baring (1988). The classical description is not a
bad approximation as long as the Landau state number of the pairs is
large (e.g. $>20$), which means that critical SR photon energy is much
larger than $\Delta E_L$, and the field strength is not too high,
i.e., $B'\leq 0.1$ (see numerical results in Harding \& Preece 1987, 
note this is the same condition for the asymptotic formula (\ref{l6})  
to hold). Both conditions are fulfilled in most cases discussed in this
paper (see Sect.\ref{HM98}).
 
For the ICS branch filial-generation, the typical energy of the
emitting (scattering) photon ($(i+1)$-th generation) is determined by
the Lorentz factor of the pairs ($(i+1)$-th generation) after they
have emitted their perpendicular energy via SR. Suppose such Lorentz
factor is $\gamma_{i+1,\parallel}$, the typical ICS photon energy is
simply (e.g. Zhang et al. 1997a)
\begin{equation} 
\epsilon_{i+1,{\rm ICS}}=2\gamma_{i+1,\parallel} B'_e 
\label{ICS} 
\end{equation} 
for resonant scatterings. Note essentially, this typical energy
depends on the local field strength, but does not depend on the
incident angle of the scatterings. This feature can be understood as
follows: a soft photon with frequency $\epsilon_0$ which impacts by an
incident angle of $\theta=\cos^{-1}
\mu$ with an electron with energy $\gamma_\parallel$, will be inverse 
Compton scattered to a maximum frequency of $\epsilon_s=2 
\gamma_\parallel^2 (1-\beta_\parallel \mu)\epsilon_0$. For a resonant 
scattering occuring, the incident energy $\epsilon_0$ should be 
Lorentz-boosted to exactly the cyclotron energy $\epsilon_B=B'$ in the 
electron rest frame, i.e., $\epsilon_0=B'/[\gamma_\parallel 
(1-\beta_\parallel \mu)]$.
Thus the angular factors in both expressions are exactly canceled 
in the final result. In other words, the electron can ``pick up'' 
the right energies of the photons from different directions and 
resonant scatter them to contribute to the same characteristic 
energy (\ref{ICS}).
 
Another point is that in (\ref{ICS}), we have again overlooked the
possible variation of $B_e$ for different generations. (We did not
specify $B_{e,i}$ for a certain generation $i$.) We will adopt
this assumption throughout the paper, since it is sound only except
for the very last generation or when resonant scattering condition
(\ref{res}) fails so that the ICS mean free path (\ref{lambdares})
becomes long. Even so, the correction to our results is still tiny.
 
The parallel Lorentz factor of the $(i+1)$-th generation pairs is (see
Appendix A, also WSL97)
\begin{equation} 
\gamma_{i+1,\parallel} ={\gamma_{i+1}\over[1+(\gamma_{i+1}^2-1)\sin^2 
\theta_{\rm kB}]^{1/2}}.  
\label{gammaparallel} 
\end{equation} 
Noting that $\gamma_{i+1}=\epsilon_i/2$ (each of the $(i+1)$-th pair
gain one half of the $i$-th generation photon energy), as long as
$\gamma_{i+1}
\gg 1$ (this is the case for most generations), we have 
$(\gamma_{i+1}^2-1)\sin^2\theta_{\rm kB} \simeq (\epsilon_i/2 
\sin\theta_{\rm kB})^2=(\chi/B'_e)^2$, which only depends on $B_e$
when we adopt a constant $\chi$. Define
\begin{equation} 
\eta_\parallel= {\gamma_{i,\parallel} \over \gamma_i} 
\end{equation} 
as the portion of the particle energy which goes to the ICS branch, from
(\ref{gammaparallel}) we obtain
\begin{equation} 
\eta_\parallel\simeq {1\over [(\chi/B'_e)^2+1]^{1/2}}.
\label{etaparallel} 
\end{equation} 
This is independent on the generation order $i$ if we neglect the very
small variation of $B_e$ in different generations. Note this is an
essential point, since the energy distribution between the two-branch (SR
and ICS) filial generations is settled for a given pulsar in different
generations of the cascades. Using (\ref{chipsr},\ref{chimspsr}), we
get
\begin{equation} 
\eta_\parallel ({\rm nPSR}) \simeq {1\over (13.5B_{e,12}^{-2}+1)^{1/2}} 
\label{etapsr} 
\end{equation} 
for normal pulsars, and
\begin{equation} 
\eta_{\parallel}({\rm msPSR}) \simeq 1.8\times 10^{-4} B_{e,9} 
\label{etamspsr} 
\end{equation} 
for millisecond pulsars. We see only a small portion of energy is left
in the ICS branch for millisecond pulsars, since $B_e$ is much
lower. This is the third factor to reduce the importance of the
ICS-branches for the millisecond pulsars. Recall the other two factors
are: longer mean free path (eq.[\ref{lambdares}]) and possible failure
of the second resonant condition (eq.[\ref{msres}]). These are the
reasons why the non-thermal X-rays are usually not important compared
with the thermal emission for millisecond pulsars, as will be shown in
Sect.\ref{nonth-x}.
 
We can also define the energy portion going to the SR branch. This is
just
\begin{equation} 
\eta_\perp=1-\eta_\parallel, 
\label{etaperp}
\end{equation} 
and was denoted as ``energy conversion efficiency'' in
WSL97. Similarly, we can derive different $\eta_{\perp}({\rm nPSR})$
and $\eta_{\perp}({\rm msPSR})$ using (\ref{etapsr}) and
(\ref{etamspsr}), respectively.
 
With (\ref{ICS}), noting $\gamma_{i+1,\parallel}=\eta_\parallel
\gamma_{i+1}$, and again $\gamma_{i+1}=\epsilon_i/2$, we have
\begin{equation} 
\kappa_{\rm ICS}={\epsilon_{i+1,{\rm ICS}} \over \epsilon_i}=
\eta_\parallel B'_e, 
\end{equation} 
or
\begin{equation} 
\kappa_{\rm ICS}({\rm nPSR})= {0.0227 B_{e,12}\over (13.5 
B_{e,12}^{-2}+1)^{1/2}} \simeq{1\over 160}B_{e,12}^2 
\end{equation} 
for normal pulsars (assuming $B_{e,12}\leq 1$) and
\begin{equation} 
\kappa_{\rm ICS}({\rm msPSR})\simeq 4.1\times 10^{-9} B_{e,9}^2 
\label{kappaicsmspsr} 
\end{equation} 
for millisecond pulsars. This means that, except for pulsars with
high magnetic fields, ICS branches usually have a more
significant decrease of the typical photon energy than SR branches,
especially for millisecond pulsars.
 
\subsection{Generation order parameters \label{zeta}} 
The so-called ``generation order parameter'' was first introduced by
Lu, Wei \& Song (LWS94) and improved by Wei, Song \& Lu (WSL97). In
their work, the authors introduced the non-integer generation order
parameter $\zeta$ to describe the conventional curvature-synchrotron
cascades based on the RS75 acceleration model, and showed that the
observed $\gamma$-ray pulsars tend to have relative large $\zeta$s.
Such a concept can delineate the sketch of the cascade process, and, 
to some extent, compensate the shortcoming of the analytic 
``recursion'' descriptions, i.e., the ignorance of the
detailed spectral features of different generations.
 
In our full-cascade picture, the complicated two-branch
filial-generation feature makes it impossible to describe the whole
process with one single generation order parameter. But we can still
borrow their definitions in our studies.
 
Suppose the primary generation $\gamma$-rays have a typical energy
$E_0$ (this is model-dependent, we'll give the expressions in the 
HM98 model (eqs.[\ref{E0-I},\ref{E0-II}]) in Sect.\ref{HM98}), 
there will be altogether
\begin{equation} 
\zeta_{\rm SR}={\log(E_{\rm esc}/E_0) \over \log (\kappa_{\rm SR})}+1 
\label{zetasr} 
\end{equation} 
SR generations before the photon energies are reduced to $E_{\rm
esc}$. Note the term $(+1)$ in (\ref{zetasr}) is to ensure $\zeta=1$
for the primary $\gamma$-rays (with energy $E_0$). $\zeta_{\rm SR}$ is
just the generation order parameter defined by LWS94 and WSL97. In our
full cascade scenario, this parameter only describe the typical
cascade branches with pure SR generations. Similarly, we can define
another parameter
\begin{equation} 
\zeta_{\rm ICS}={\log(E_{\rm esc}/E_0) \over \log (\kappa_{\rm ICS})}+1 
\label{zetaics} 
\end{equation} 
to describe the pure ICS cascade branches. However, neither
(\ref{zetasr}) nor (\ref{zetaics}) can describe the ``crossed''
generations (e.g. ICS branch from the SR parent generation or vice
versa). We will define some more general generation order parameters
(e.g. [\ref{zetaicsk}],[\ref{zetac}]) to construct an analytic
estimate formula
of pulsar X-ray luminosities (see eqs.[\ref{Lxi},\ref{Lxni}]).
For most pulsars, the ICS branch photons usually do not pair-produce
any more (see Sect.\ref{HM98}, also SDM95). This can further
simplify the description (see details in Sect.\ref{nonth-x}).
 
\subsection{HM98 acceleration model \label{HM98}} 
Here before, we have not specified an acceleration model. The
descriptions above are valid for any polar cap acceleration model once
$E_0$ is specified.  There are generally two subclasses of polar cap
acceleration models. One subclass is the vacuum gap model first 
proposed by RS75, and modified by Usov \& Melrose (1995; 1996) and 
Zhang et al. (Zhang \& Qiao 1996; Zhang et al. 1997a,b).
Another subtype model is the space-charge-limited flow model
(Arons \& Scharlemann 1979; Arons 1983), which assumes free flow of
particles from the surface. Such a model is improved by Muslimov \&
Tsygan (1992) and Muslimov \& Harding (1997) by incorporating the
frame dragging parallel electric fields, and finally HM98
imposed the zero $E_\parallel$ boundary condition at the pair
formation front (PFF) to treat the accelerator more realistically.
In reality, the environment of
different pulsars could vary much from one to another, so that the
above-mentioned two different types of inner accelerators may anchor 
in different pulsars. For example, there is growing evidence showing
that a handful of the ``drifting'' radio pulsars seem to have the 
RS-type vaccum accelerators in their polar cap region (Deshpande 
\& Rankin 1999; Vivekanand \& Joshi 1999). However, theoretical 
arguments, i.e., the so-called binding energy problem encountered
by the RS75 model (Jones 1985, 1986; Neuhauser et al. 1986, 1987),
support the space-charge-limited flow acceleration scheme.
In this paper, we will adopt the HM98 acceleration model 
uniformly for all the pulsars.

No simple analytic description for all pulsars is available in the
HM98 model due both to the complicated form of $E_\parallel$ within 
the gap and to the screening details. Nevertheless, a general
accelerating picture is available, i.e., $E_\parallel$ grows
approximately linearly if the gap length $S_c$ is smaller than the
``effective'' polar cap radius $r_{pc,E}$ (assuming an effective
star surface $R_E\sim (1.5-2) R$), or will be saturated to an almost
constant value above $r_{pc,E}$ if $S_c > r_{pc, E}$. 
Hereafter we will define the cases with or without field saturation 
as the regime I and II, respectively. In principle, both CR and ICS
of the primary particles should be taken into account. However, for
most pulsars (except for high $B$ pulsars), as pointed out by HM98,
as long as ICS dominates in the accelerator, the accelerator itself is
unstable, since the upward ICS occurs in the resonant regime, while
the downward ICS occurs in the Klein-Nishina regime. Thus the final
stable accelerator should be controlled by CR. In fact, the ``effective 
surface'' $R_E$ is determined by the criterion that beyond this height, 
the ICS energy loss rate is less than the CR energy loss rate (see HM98). 
In regime I, the CR-controlled gap length ($S_c$), the maximum typical 
Lorentz factor of the primary particles ($\gamma_0$), as well as the 
characteristic CR photon energy emitted by the primary particles, i.e., 
$E_0={3\over 2}{\gamma^3\hbar c\over \rho_{e}}$, could be approximately 
obtained using (A3) of HM98, which read (see also eqs.[77,79] in HM98, 
in which there are some minor computing errors)
\begin{equation} 
S_c({\rm I}) \simeq 4.2\times 10^4 {\rm cm}
B_{p,12}^{-4/7}P^{4/7} R_{E,6}^{16/7} (\cos\alpha)^{-3/7},
\label{Sc-I} 
\end{equation} 
\begin{equation} 
\gamma_0({\rm I})\simeq 8.2\times 10^7 B_{p,12}^{-1/7} P^{1/7} 
R_{E,6}^{4/7} (\cos\alpha)^{1/7},
\label{gamma0-I} 
\end{equation}
and
\begin{equation} 
E_0({\rm I}) \simeq 178 ({\rm GeV}) B_{p,12}^{-3/7}P^{-1/14}R_{E,6}
^{17/14}(\cos \alpha)^{3/7},
\label{E0-I} 
\end{equation}
respectively, where $\alpha$ is the inclination angle of the neutron 
star. For the regime II, we adopt the saturated $E_\parallel$ 
presented in (A5) of HM98, and obtain the different expressions 
which read
\begin{equation} 
S_c({\rm II}) \simeq 3.2\times 10^6 {\rm cm}
B_{p,12}^{-1}P^{7/4} R_{E,6}^{-2} (\cos\alpha)^{-3/4},
\label{Sc-II} 
\end{equation} 
\begin{equation} 
\gamma_0({\rm II})\simeq 1.4\times 10^7 P^{-1/4} R_{E,6}^2 
(\cos\alpha)^{1/4},
\label{gamma0-II} 
\end{equation}
and
\begin{equation} 
E_0({\rm II})\simeq 0.88 ({\rm GeV}) P^{-5/4}R_{E,6}
^{11/2}(\cos \alpha)^{3/4}.
\label{E0-II} 
\end{equation}
The emission height discussed in the previous sections is then
\begin{equation}
r_{e,6}=R_{E,6}+S_{c,6}.
\label{re6-RE6}
\end{equation}
It is worth noting that, the primary particles actually lose
their energies within and above the accelerator via both CR 
and ICS, and the ICS process has a different typical $E_0$, so 
that it will also contribute to a component in the final 
$\gamma$-ray spectrum and change the generation structure to some
extent. But such modifications are not prominant. Furthermore, the 
energetics of both the CR and ICS photons are from the kinetic energy 
of the primary particles (\ref{gamma0-I},\ref{gamma0-II}), thus the 
variation of the detailed primary photon spectrum does not change much 
the luminosity predictions presented in Sect.3, where we adopt the 
primary luminosity as the total polar cap particle luminosity (see 
[\ref{Lpc}]).

Since regime I is limited by regime II, in principle, we should adopt
the minimum of $\gamma_0$(I) and $\gamma_0$(II) as the real $\gamma_0$
(and hence $E_0$) for a certain pulsar. The critical condition to 
separate the two regimes could be obtained by equating 
eq.(\ref{gamma0-I}) with (\ref{gamma0-II}). Thus a pulsar should be
in regime I if
\begin{equation}
B_{p,12}^{1/7}P^{-11/28}R_{E,6}^{10/7}(\cos\alpha)^{3/28}>6.0,
\label{boundary}
\end{equation}
and in regime II otherwise. Note that this is just a rough analytic
treatment. Numerically, there should be no such distinct boundary,
and the two regimes should meet smoothly.

With (\ref{E0-I}) or (\ref{E0-II}), one can get explicit expressions
of the generation parameters from (\ref{zetasr}) and (\ref{zetaics}).
This completes the analytic description of the full cascade process.
It is worth noting that $1<\zeta_{\rm ICS}<2$ for most pulsars
(see Table 2). Such an issue was also noted by SDM95, who thus
suggested that ICS branches will not further produce filial-generation 
pairs. However, for the high field pulsars, this is not necessarily
the case (also see Table 2). In our analytic formula
below, without losing generality, we will also cover the possible
contributions from the ICS-photon-produced pairs (see eq.[\ref{Lxi}]).
 
HM98 has discussed the positron backflow feedback quantitatively. 
Though the complication of the $E_\parallel$ screening
process prevents an accurate determination of the fraction of
backflowing positrons, an upper limit of such a flow is readily
available from the model, which is useful to estimate the maximum
thermal X-ray luminosities due to polar cap heating. The expressions
of such maximum thermal X-ray luminosities are shown in equations
(\ref{Lx-th-I}) and (\ref{Lx-th-II}) (see Sect.\ref{th-x}).
 
\section{Luminosity predictions} 
\subsection{Gamma-ray luminosity} 
Harding (1981) pointed out that by assuming a roughly constant
potential in the inner accelerators for different pulsars (so that the
primary particles gain almost the same energy $\gamma_0$ for different
pulsars), the total polar cap ``luminosity''
\begin{equation} 
L_{\rm pc}=\gamma_0 mc^2 \dot N_{ p}
\label{Lpc} 
\end{equation}  
is then proportional to $B_p/P^2$, since the particle flow rate
\begin{equation} 
\dot N_p=c n_{_{\rm GJ}} \pi r_p^2=1.4 \times 10^{30} R_6^3 B_{p,12} 
P^{-2} (\cos\alpha)
\label{Np} 
\end{equation} 
itself is proportional to $B_p/P^2$, where $n_{_{\rm GJ}}\simeq {{\bf
\Omega\cdot B}\over 2\pi e c}$ is the Goldreich-Julian (1969) number
density. This will give a natural interpretation of the observed
relationship of $L_\gamma\propto (L_{\rm sd})^{1/2}$ if the polar 
cap ``luminosity'' is completely converted to $\gamma$-ray emission.  
However, in the canonical CR-SR (DH96) or ICS-SR (SDM95) cascade 
picture, the expected $\gamma$-ray luminosity $L_\gamma$ is not 
fully identical to $L_{\rm pc}$, since only the perpendicular 
portion of the energies of the pairs goes to the next generation 
radiation. More specifically, one has (WSL97)
\begin{equation} 
L_\gamma({\rm canonical})\simeq L_{\rm pc}\eta_\perp^{\zeta_{\rm SR}},
\label{Lgammaold} 
\end{equation} 
which will deviate from $L_{\rm pc}$ to some extent. The deviation is
more significant for the pulsars with small $\eta_\perp$ and large
$\zeta_{\rm SR}$ (e.g. PSR 1509-58, see Fig.3, Table 1). In the full 
two-branch cascade picture described above, it is safe to regard
\begin{equation} 
L_\gamma({\rm full})\simeq L_{\rm pc},
\label{Lgamma} 
\end{equation} 
since the ICS branches ``pick up'' the ``lost'' parallel kinetic
energies of the particles also to radiation, so that almost 100\% 
of the polar cap ``luminosity'' is converted to high energy
radiation. Although not all these energies are converted to
$\gamma$-rays (in fact, the non-thermal X-ray emission is also
part of this energy budget, see Sect.\ref{nonth-x} below), the
majority of this energy output is in the $\gamma$-ray band.

In the HM98 model, the typical energy of the primary particles
$\gamma_0$ turns out to be very weakly dependent on the pulsar
parameters (eqs.[\ref{gamma0-I},\ref{gamma0-II}]), especially
for the young pulsars in regime I (only Geminga and PSR 1055-52
are in regime II). This is in agreement with the assumption of 
Harding (1981). With eqs.(\ref{Lgamma},\ref{Lpc},\ref{Np}) and
(\ref{gamma0-I}) or (\ref{gamma0-II}), and
\begin{equation}
L_{\rm sd}=-I\Omega \dot\Omega \simeq 9.68\times 10^{30} 
B_{p,12}^2 P^{-4}I_{45}
\label{Lsd}
\end{equation}
($I=10^{45}I_{45}$ is the moment of inertia), the $\gamma$-ray 
luminosity is finally
\[ 
L_\gamma({\rm full, I})\simeq 9.4\times 10^{31} {\rm erg\cdot
s^{-1}}B_{p,12}^{6/7} P^{-13/7}R_{E,6}^{4/7}(\cos\alpha)^{8/7}
\] 
\begin{equation} 
=3.0\times 10^{16} B_{p,12}^{-1/7}P^{1/7}R_{E,6}^{4/7}
(\cos\alpha)^{8/7}(L_{\rm sd})^{1/2},
\label{Lgamma-I}
\end{equation} 
for regime I, and
\[ 
L_\gamma({\rm full, II})\simeq 1.6\times 10^{31} {\rm erg\cdot
s^{-1}}B_{p,12} P^{-9/4}R_{E,6}^2(\cos\alpha)^{5/4}
\] 
\begin{equation} 
=0.5\times 10^{16} P^{-1/4}R_{E,6}^2
(\cos\alpha)^{5/4}(L_{\rm sd})^{1/2},
\label{Lgamma-II}
\end{equation}
for regime II. This almost reproduces the $L_\gamma\propto 
(L_{\rm sd})^{1/2}$ feature (also see Fig.3). 

A natural implication of the $L_{\rm pc} \propto (L_{\rm sd})
^{1/2}$ dependence is that the $\gamma$-ray emission efficiency, 
$L_{\rm pc}/L_{\rm sd}$, will increase for the pulsars with 
lower $L_{\rm sd}$. Physically, such an efficiency can not be 
greater than unity. Since as $L_{\rm pc}$ approaches 
$L_{\rm sd}$ the accelerator should be in regime II (saturated),
we can use (\ref{Lgamma-II}) and (\ref{Lsd}) to get the constraint
\begin{equation}
B_{p,12}P^{-7/4} R_{E,6}^{-2}(\cos\alpha)^{-5/4} \geq 1.65.
\label{constraint}
\end{equation}
Such a line is also close to the pulsar pair formation 
``deathline'' (for a detailed discussion of the deathlines,
see Zhang, Harding \& Muslimov (1999)). The luminosity formulae 
(\ref{Lgamma-I},\ref{Lgamma-II}) are no longer valid when 
(\ref{constraint}) fails. None of the observed spin-powered 
X-ray pulsars studied in this paper are beyond this line.

In Table 1 and Fig.3, we compare the broad-band high energy
luminosities of the 8 $\gamma$-ray pulsars with the theoretical 
predictions both in the full cascade model and the canonical CR-SR 
cascade model. Though the $L_\gamma\propto (L_{\rm sd})^{1/2}$ feature 
is reported using $\gamma$-ray luminosities above 100KeV (Thompson et
al. 1997), here we adopt the latest luminosity data above 1eV for the
7 previously known $\gamma$-ray pulsars including the new data of PSR 
1055-52 (Thompson et al. 1999), which also clearly shows the same trend, 
and indicates that most emission is spread in the $\gamma$-ray band. 
For the newly discovered $\gamma$-ray pulsar PSR 1046-58, we adopt the
luminosity data above 400MeV (Kaspi et al. 1999). As shown in Fig.3, all
these data just meet our model prediction $L_{\gamma}(\rm full)$ 
(eqs.[\ref{Lgamma-I},\ref{Lgamma-II}]) well, which is essentially the 
polar cap luminosity $L_{\rm pc}$ (eq.[\ref{Lpc}]). The model parameters 
we adopt are: $\alpha=30^{\rm o}$ and $R_{E,6}=1.8$. It is clear 
that the full cascade model shows a much better fit to the 
observations than the canonical model which predicts lower 
$\gamma$-ray luminosities (eq.[\ref{Lgammaold}]) (see Fig.3), 
especially for PSR 1509-58. The small differences between the 
full-cascade predictions and the observations are within the 
range of different $\alpha$ and slightly different 
$R_{E,6}$. Thus our model is in good agreement with the observations. 
Note that the present description is incomplete for the high
$B$ pulsar PSR 1509-58 (with $\log B_p=13.49$). In such strong fields, 
the resonant frequency is much higher, and the downward ICS of the positrons 
also occurs in the resonant regime (HM98). The anisotropy of ICS processes 
then dissappears and the accelerator should still be located at the neutron 
star surface, with ICS as the controlled mechanism. However, photon 
splitting will also play an important role in high $B$ pulsars (HBG97),
so that the cascade is more complicated and includes both pair production
and photon splitting. All these effects may alter the luminosity predictions 
of high $B$ pulsars significantly, both for $\gamma$-rays and also for X-rays 
as well.

\centerline{}
\centerline{\psfig{file=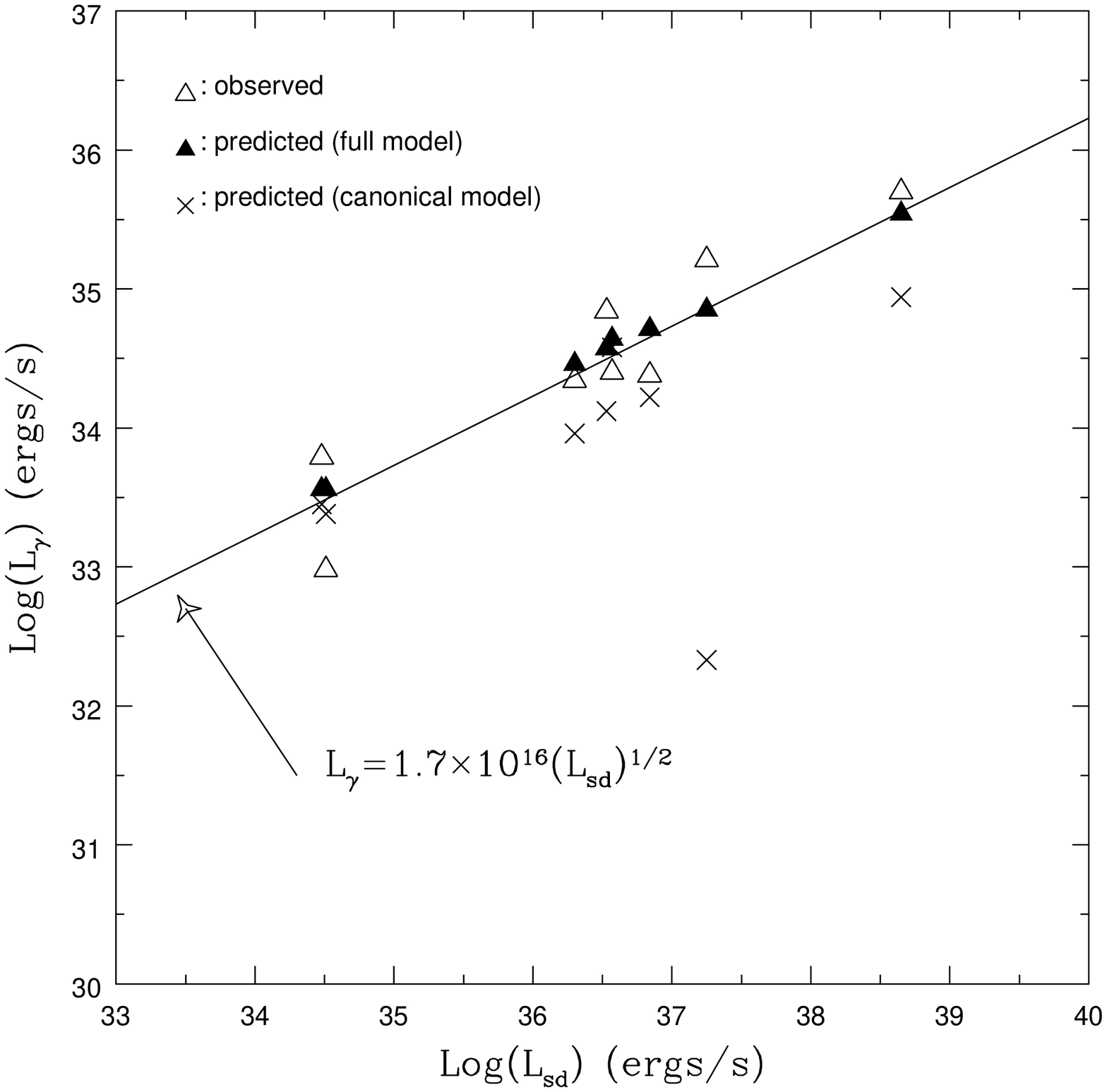,width=8.0cm}} 
\figcaption{Comparison of the observed broad-band luminosities for the 
7 $\gamma$-ray pulsars and $\gamma$-ray luminosity of PSR 1046-58 above
400MeV with the model predictions in both the full cascade and the
canonical CR(ICS)-SR cascade models. Note how the full cascade  
model `picks up' the luminosities which the canonical model  
has `lost', and clearly reproduces the $L_\gamma\propto (L_{\rm sd}) 
^{1/2}$ feature.
\label{fig:gamma}} 
\centerline{}
 
It is also of interest to examine $\gamma$-ray luminosities of the
millisecond pulsars. According to (\ref{Lgamma-II}), usually the 
millisecond pulsars also have considerable $\gamma$-ray luminosities
($\propto (L_{\rm sd})^{1/2}$). However, due to their weak field 
strengths, the escaping photon energy is quite high (eq.[\ref{esc2}]).
As a result, $\gamma$-ray spectra of the millisecond pulsars are
expected to be very hard, with the lowest characteristic SR energy 
(eq.[\ref{Ecmin}]) about 150GeV, and the detectable photon number is 
fewer. Future missions, such as GLAST, may detect emission from
these sources.

\subsection{Thermal X-ray luminosity \label{th-x}} 
 
As reviewed in Sect.\ref{full}, there are many possible mechanisms 
to account for both the full surface and the hot polar cap thermal 
emission components from neutron stars. Here we will treat the full
surface emission by adopting a simple ``standard'' cooling model,
and treat the hot spot emission using the polar cap heating scenario
predicted by the HM98 model.

\subsubsection{Full surface thermal emission: neutron star cooling}

Neutron star cooling is the results of many different mechanisms,
the physical details of some of which are still poorly known. Thus
there are many uncertainties in different models (see Shapiro
\& Teukolsky (1983) for a review). In our calculation, we use a
rough relation
\begin{equation}
\begin{array}{cc}
\log T_{s,6}=-0.1 \log \tau +0.37 & \log\tau\leq 5.2 \\
\log T_{s,6}=-0.5 \log \tau +2.45 & \log\tau > 5.2 \\
\end{array},
\label{cooling}
\end{equation}
which is obtained by a rough fit to the latest detailed numerical
model of Schaab et al. (1999, their Fig.3), and is consistent with
the standard ``modified URCA'' neutrino cooling plus photon cooling 
model reviewed in Shapiro \& Teukolsky (1983). Here, $\tau=P/2\dot P$ 
is the spin-down age of the pulsars. The reasons for us to adopt
such a simple treatment are as follows: Firstly, such simple cooling
models are not inconsistent with the observations (BT97; Schaab et al.
1999). Secondly, as long as the resonant ICS condition (\ref{res})
is satisfied, our non-thermal luminosity predictions 
are insensative to the temperature of the soft photons. Thus some
extent of inaccuracy in the full-surface temperature adoption has 
little influence on the final results. Thirdly, in the polar cap
cascade scenarios, it is unlikely to form a nearly static pair
``blanket'' near the surface, as assumed in the outer gap models. 
Hence, there should be no ``reflected'' full-surface thermal
emission component. Furthermore, the outer gap models did not 
compare their soft thermal emission prediction with the observations,
so that the assumption can not be justified.

\subsubsection{Polar cap heating}

As discussed in Sect.\ref{HM98}, the complication of the $E_\parallel$
screening process makes it difficult to treat the positron backflow 
accurately, so that an explicit analytic description of the polar cap 
heating is not available. In principle, the maximum backflow positron 
rate $\dot N_{e^+,{\rm max}}$ could be estimated by multiplying the 
polar cap current $\dot N_p$ (eq.[\ref{Np}]) with a factor
\begin{equation}
f\simeq \left| {(\nabla \cdot {\bf E})_\parallel/(4\pi) \over
2\rho_{_{GJ}}} \right|,
\label{f}
\end{equation}
so that\footnote{Note that this estimate of $L_{e^+,{\rm max}}$ 
is much smaller than that given in eqs.(55,59) of HM98 because the 
screened accelerator length has been included here.}
\begin{equation} 
L_{e^+,{\rm max}}=\gamma_0 mc^2 \dot N_{e^+,{\rm max}}=f L_{\rm pc},
\label{Le+max}
\end{equation}  
since $E_\parallel$ determines the charge density required for sceening, 
and the factor (1/2) accounts for the readjustment of the primary 
current due to the backflow current. However, due to the mathematical 
complications of the space-charge-limited flow model with frame-dragging
included (Muslimov \& Tsygan 1992; Muslimov \& Harding 1997; HM98),
a simple general expression is not available. For the unsaturated 
case, the gap is pancake-shaped, and the boundary conditions at the
open field lines are not important. Thus one has $(\nabla \cdot 
{\bf E})_\parallel/(4\pi)\simeq (\nabla \cdot {\bf E})/(4\pi)$, 
which is essentially $(\rho-\rho_{_{GJ}})$, so that
\[ 
f({\rm I})\simeq \left|{\rho-\rho_{_{GJ}}\over 2\rho_{_{GJ}}}\right|
\simeq{\eta_*^2\kappa(1-\eta^{-3}) \over (1-\eta_*^2\kappa \eta^{-3})}
\]
\begin{equation}
={0.075 (R_{E,6}^{-3}-r_{e,6}^{-3}) \over 1-0.15r_{e,6}^{-3}},
\label{fI}
\end{equation}
where eqs.(10,11) of HM98 has been used, $(\sin\alpha)$ term is 
neglected, and $\eta_*=R/R_E=R_{E,6}^{-1}$, $\eta=r/R_E$, $\kappa=
(r_g/R_E)(I/MR^2)=\kappa_0 R_{E,6}^{-1}$, $\kappa_0=(r_g/R)(I/MR^2)
\sim 0.15$ have been adopted. Since usually $S_c({\rm I}) \ll R$,
with (\ref{re6-RE6}), one can further derive
\begin{equation}
f({\rm I})\simeq 0.23 R_{E,6}^{-3}z,
\label{f-I}
\end{equation}
where 
\begin{equation}
z=S_c({\rm I})/R_E=4.2\times 10^{-2}B_{p,12}^{-4/7}P^{4/7}
R_{E,6}^{9/7}(\cos\alpha)^{-3/7},
\label{z} 
\end{equation} 
and eq.(\ref{Sc-I}) has been used. 
If this backflow energy is entirely converted to thermal emission 
at the hot polar cap, then the maximum polar cap thermal X-ray 
luminosity should be
\begin{equation}
L_{x,pc,{\rm max}}\simeq L_{e^+,{\rm max}},
\label{Lxthmax}
\end{equation}
which is
\begin{equation}
L_{x,pc,{\rm max}}({\rm I})\simeq 9.1\times 10^{29}{\rm erg\cdot
s^{-1}}B_{p,12}^{2/7}P^{-9/7}R_{E,6}^{-8/7} (\cos\alpha)^{5/7},
\label{Lx-th-I} 
\end{equation} 
for regime I. The maximum polar cap temperature (assuming an area of 
$\pi r_{pc}^2$, where $r_{pc}=\theta_{\rm pc}R=1.45\times 10^4P^{-1/2}
{\rm cm}$) is hence
\[ 
T_{\rm pc,max}({\rm I})=\left({L_{x,th,{\rm max}}({\rm I})\over 
\sigma\pi r_{pc}^2}\right)^{1/4}
\] 
\begin{equation} 
\simeq 2.2 \times 10^6 {\rm K}
B_{p,12}^{1/14}P^{-1/14}r_{e,6}^{-2/7}(\cos\alpha)^{5/28},
\label{T-I} 
\end{equation} 
where $\sigma=5.67\times 10^{-5}{\rm ergs\cdot cm^{-2}\cdot
K^{-4}\cdot s^{-1}}$ is Stefan's constant. 

For regime II (saturated case), $S_c$ is larger than the
effective polar cap radius $r_{pc,E}=r_{pc}R_{E,6}^{3/2}$.
The gap shape becomes narrow and long. The boundary conditions
at the open field line boundaries become important, so that
$(\nabla \cdot {\bf E})_\parallel/(4\pi)$ can be much
less than $(\nabla \cdot {\bf E})/(4\pi)$. The 
approximation (\ref{fI}) no longer holds. To achieve a very 
accurate description of such reversed positron fraction, one 
has to appeal to complicated mathematical descriptions 
(e.g. Muslimov \& Tsygan 1992; HM98) or numerical simulations
to get $(\nabla \cdot {\bf E})_\parallel/(4\pi)$. However, 
noticing that $(\nabla \cdot {\bf E})_\parallel/(4\pi)
\simeq (\nabla \cdot {\bf E}_\parallel)/(4\pi)\simeq \partial
E_\parallel/\partial r$, we can still get an approximate
formula for $f$ by adopting (A4) of HM98, and making some
reasonable simplication, which finally reads
\[
f({\rm II})\simeq 3\kappa_0(1-\xi^2)\left[{f(\eta)\over 
f(\eta_{*})}\right]^{-3}(1-\eta_{*}^2{\kappa\over\eta^3})^{-1}
r_{e,6}^{-2}R_{E,6}^3\left({\Omega R\over cf(1)}\right)
\]
\begin{equation}
\simeq 5.7\times 10^{-5}r_{e,6}^{-2}R_{E,6}^3 P^{-1},
\label{f-II}
\end{equation}
where the function $f(\eta)$ is the correction factor of the
polar cap radius in curved spacetime (see eq.[8] of HM98), 
which is equivalent to the function $q(r)$ introduced in this
paper (see eq.[\ref{q}]). Other notions have been described
above, and typical values have been adopted.
With eqs.[\ref{Lxthmax},\ref{Le+max},\ref{Lgamma-II},\ref{f-II}], 
it is easy to estimate the polar cap heating luminosity in 
regime II,
\begin{equation}
L_{x,pc,{\rm max}}({\rm II})\simeq 9.1\times 10^{26}{\rm erg\cdot
s^{-1}}B_{p,12}P^{-13/4}r_{e,6}^{-2}R_{E,6}^5 (\cos\alpha)^{5/4},
\label{Lx-th-II} 
\end{equation} 
and the polar cap temperature
\begin{equation} 
T_{\rm pc,max}({\rm II})\simeq
0.53 \times 10^6 {\rm K}B_{p,12}^{1/4}P^{-9/16}
r_{e,6}^{-1/2}R_{E,6}^{5/4}(\cos\alpha)^{5/16}.
\label{T-II} 
\end{equation} 
We see that $T_{\rm pc,max}({\rm II})$ is sensitive to pulsar
period ($\propto P^{-9/16}$) while $T_{\rm pc,max}({\rm I})$ 
is not ($\propto P^{-1/14}$). This is just the refection of
the open field line boundary conditions on the saturated gap.
We also notice that the peak energy of such thermal emission 
falls into the ROSAT band. However, it is notable that the 
thermal X-ray luminosity (eqs.[\ref{Lx-th-I}] or [\ref{Lx-th-II}]) 
does not have the dependence that is observed in ROSAT band 
($L_{\rm x}(ROSAT)\propto L_{\rm sd}$, BT97). This rules out the 
possibility of interpreting ROSAT luminosities in terms of 
only thermal emission due to polar cap heating. We will show next, 
however, that a combination of such thermal luminosities with the 
non-thermal luminosities due to the ICS of higher generation pairs 
could roughly reproduce such dependence. The thermal component 
discussed above will dominate the ROSAT band emission in most of 
the millisecond pulsars. Another point is that, the luminosity and
polar cap temperature formulae presented above 
(eqs.[\ref{Lx-th-I},\ref{T-I},\ref{Lx-th-II},\ref{T-II}) are still
upper limits, especially for the saturated regime (regime II), 
though actual values are not much lower. For 
the millisecond pulsars which are near the deathlines (Zhang et al.
1999), such deviations could be even larger, since our approximations
adopted to describe the accelerator are no longer accurate enough.

\subsection{Non-thermal X-ray luminosity \label{nonth-x}} 
In our full-cascade picture, we can naturally get a nonthermal X-ray
emission component from the soft tail of the ICS spectra of the higher
generation pairs. Attaining an accurate spectrum requires a more
careful modeling using Monte Carlo simulation approach, which will
not be done in this paper. However, some simple estimate of the
non-thermal X-ray luminosities of pulsars could be achieved using the
analytic method described in Sect.\ref{gop} and \ref{zeta}.
 
Let us compute the luminosity below a certain energy
$E_c$. In principle, one should add the contributions from all the
branches over the whole cascades. However, it is notable that the SR
spectra of all the SR branches can not get down to as low energy as
the X-rays observed by ROSAT and ASCA (Harding \& Daugherty 1999;
Rudak \& Dyks 1999). This could be justified as follows. The lowest 
characteristic SR energy in the full-cascade is
\begin{equation} 
E_{\rm c,min}({\rm SR})=E_0 {\kappa_{\rm SR}}^{\zeta_{\rm SR}},
\label{Ecmin}
\end{equation} 
and in principle, the minimum SR energy is the blueshifted cyclotron
energy (corresponding to the electron occupying only one Landau state,
see Harding \& Daugherty 1999)
\begin{equation} 
E_{\rm min}({\rm SR})=\gamma_{\zeta_{\rm
SR},\parallel}\hbar\omega_{B,e}=
\gamma_{\zeta_{\rm SR},\parallel}mc^2 B'_e, 
\label{SRmin} 
\end{equation} 
where
\begin{equation} 
\gamma_{\zeta_{\rm SR},\parallel}={\epsilon_0 {\kappa_{\rm SR}}
^{(\zeta_{\rm SR}-1)} \over 2}\eta_\parallel 
\end{equation} 
is the initial parallel energy of the last generation ($\zeta_{\rm
SR}$-th generation) pairs.
 
For a typical normal pulsar with $P=0.1$s, $B_{p,12}=5.8$ (thus
$B_{e,12}\simeq B_{p,12}R_{E,6}^{-3}=1$ for $R_{E,6}=1.8$), and 
$\alpha=30^{\rm o}$, we have $E_0\sim 196$GeV
(eq.[\ref{E0-I}]), $E_{\rm esc}=0.52$GeV (eq.[\ref{esc1}]), $\kappa_{\rm
SR}\sim 1/16$ (eq.[\ref{kappasr1}]), $\zeta_{\rm SR}=3.14$
(eq.[\ref{zetasr}]), and $\eta_\parallel =0.237$ (eq.[\ref{etapsr}]),
so that $E_{\rm c,min}({\rm SR,nPSR})\sim 32.8$MeV and $E_{\rm
min}({\rm SR,nPSR})=1.26$MeV. Both values are much higher than the
X-ray bands we are interested in (e.g. below 10keV). Similarly, for a
typical millisecond pulsar with $P\sim 0.005$s, $B_{p,9}=0.5$,
$R_{E,6}=1.2$, and $\alpha=30^{\rm 0}$, we have $E_0\sim 163$GeV
(eq.[\ref{E0-II}]), $E_{\rm esc}=159$GeV (eq.[\ref{esc2}]), $\kappa_{\rm
SR}\sim 1/11$ (eq.[\ref{kappasr2}]), $\zeta_{\rm SR}=1.01$
(eq.[\ref{zetasr}]), and $\eta_\parallel=1.6\times 10^{-5}$
(eq.[\ref{etamspsr}]), so that $E_{\rm c,min}({\rm SR,msPSR})\sim
149$GeV and $E_{\rm min}({\rm SR,msPSR})=0.026$KeV. Note although the
lowest end of this SR spectra could be down to as low as 0.026KeV,
there is a very wide span of energy of the last SR generation
starting from more than 100 GeV. Thus the SR contribution to the ASCA 
and ROSAT band X-rays is tiny compared to the ICS contributions 
as described below.
 
For the ICS branches, the expression of characteristic energy
(eq.[\ref{ICS}]) is similar to the minimum SR energy
(eq.[\ref{SRmin}]), but the physical meaning is different (see
Sect.\ref{gop}). The most important difference is, as long as the
resonant scattering condition (eq.[\ref{res}]) is satisfied, the
Lorentz factor of the pairs will keep decreasing so that the
characteristic energy of the ICS process will cover a broad energy
range, the low energy end of which could present a significant
contribution to the X-ray band we are interested in.

As described above, the complication of the full-cascade picture
makes it difficult to obtain an analytic expression which includes
all the possible contributions to the X-ray luminosity. The main
complication arises from the ``mixed'' generations. However,
there are two essential features that make such an analytic
expression possible: (1) The energy distribution to the two
branches is fixed for a certain pulsar (e.g. $\eta_\perp$ and
$\eta_\parallel$ are constant for a certain pulsar regardless
the generation order, see eqs.[\ref{etaparallel},\ref{etaperp}]) 
if we ignore the small variation of $B_e$ in the different 
generations; (2) Only ICS branches can have significant
contributions to X-ray luminosities. In principle, The upper
limit of the non-thermal X-ray luminosity below a certain energy
$E_c$ could be estimated as
\begin{equation} 
L_{x,nth}(E_c)\lesssim L_{\rm pc}\sum_{k=1}^{\rm int(\zeta_{\rm
SR})} \left[\eta_\perp^{k-1} \left(\sum_{j=1}^{\rm 
int(\zeta_{\rm ICS,k-1})}\eta_\parallel^{j} \eta_{c,k,j} 
\right)\right].
\label{Lxi} 
\end{equation} 
Here $\zeta_{\rm SR}$ is just the pure SR generation order parameter
defined in (\ref{zetasr}), and $\zeta_{\rm ICS,k}$ is some more
general ICS generation order parameter defined by
\begin{equation}
\zeta_{\rm ICS,k}={\log(E_{\rm esc}/E_k) \over
\log(\kappa_{\rm ICS})}+1,
\label{zetaicsk}
\end{equation}
which is the number of pure ICS generations for the typical energy
of the $k$-th SR generation,
\begin{equation} 
E_k=E_0 \kappa_{\rm SR}^{k},
\label{Ek}
\end{equation} 
to reduce to the escaping energy $E_{\rm esc}$ (eqs.[\ref{esc1},
\ref{esc2}]). Note that we have defined that, for $k=0$,
$E_k$ is just $E_0$ (eqs.[\ref{E0-I}] or
[\ref{E0-II}]). To keep consistency with
(\ref{zetasr}) and (\ref{zetaics}), we still keep the term (+1) in
(\ref{zetaicsk}), but use $(k-1)$ in (\ref{Lxi}) and (\ref{gammakj})
to subtract this extra generation.
Another important parameter in (\ref{Lxi}), $\eta_{c,k,j}$, is to
describe the portion of the particles' energy that goes to the band
below certain $E_c$ in a certain ICS branch, which is defined as the
ratio of the Lorentz factor required to produce ICS photons with
energy $E_c$,
\begin{equation} 
\gamma_c=E_c/[(1-\beta\mu)2.8 kT], 
\label{gammac}
\end{equation} 
to the initial parallel Lorentz factor of particles from the
$(j-1)$-th ICS-photon-produced pairs from the $(k-1)$-th
SR-photon-produced pairs,
\begin{equation} 
\gamma_{k,j}={\epsilon_0\kappa_{\rm SR}^{k-1} \kappa_{\rm ICS}^{j-1}
 \over 2} \eta_\parallel, 
\label{gammakj}
\end{equation} 
so that
\begin{equation} 
\eta_{c,k,j}=\left\{ 
 \begin{array}{ll} 1, & \gamma_c \geq \gamma_{k,j} \\
 \gamma_c/\gamma_{k,j}, & \gamma_c < \gamma_{k,j}
 \end{array} \right. ,
\label{etackj} 
\end{equation} 
and the truncation of $\eta_{c,k,j}=1$ at
$\gamma_c\geq \gamma_{k,j}$ ensures the
energy portion can not exceed 100\%.
Note that (\ref{Lxi}) has included most of the branches that have
contributions to the X-ray luminosity. There are still some branches
missing, e.g., the ICS branch from the SR branch produced by the 
even earlier ICS branch. In any case, the correction of
these ``higher order'' branches to (\ref{Lxi}) is tiny.

The estimate in (\ref{Lxi}) actually involves the assumption that
the conversion efficiency from particle kinetic energy to radiation
in the ICS branches is 100\%.
This is only true when the resonant scattering condition (\ref{res})
is fulfilled. If the resonant condition fails, the estimate in
(\ref{Lxi}) is then just an upper limit. Thus it is essential to
compare $\gamma_{\rm res}$ (eq.[\ref{res}]) with $\gamma_c$
(eq.[\ref{gammac}]). If $\gamma_{\rm res}\ll \gamma_c$, the ICS
processes producing photons with $E_c$ are well
within the resonant regime, then $\eta_{c,k,j}$ is essentially the
precise energy portion. If $\gamma_{\rm res}\sim \gamma_c$, the ICS
processes can still produce photons with $E_c$. But the process just
marginally occurs in the resonant regime, thus $\eta_{c,k,j}$ is a
marginal upper limit. If, however, $\gamma_{\rm res}\gg \gamma_c$, the
resonant scattering has stopped at a much higher energy than
$\gamma_c$. Then only a small portion of the photons with $E_c$ could
be produced, then $\eta_{c,k,j}$ could just be regarded as a very loose
upper limit. Such differences are reflected in Table 3. Note that 
when calculating the X-ray luminosity within a certain band, $E_{c1}
<E<E_{c2}$, we have actually taken $L_{\Delta E}=L_{x,nth}(E_{c2})-
L_{x,nth}(E_{c1})$ as the predicted luminosity in this band. In Table
3, we have adopted `$\sim$', `$\lesssim$', and `$<$', for the cases
of $\gamma_{\rm res}\leq \gamma_{c1}$, $\gamma_{c1}\leq\gamma_{\rm res}
\leq\gamma_{c2}$, and $\gamma_{\rm res}>\gamma_{c2}$, respectively.
For old normal pulsars such as PSR 1929+10 and PSR 0950+08, since both
the cooling temperature (\ref{cooling}) and the polar cap temperature
(\ref{T-II}) are much lower, the ICS efficiency is lower. Unless
$\gamma_{\rm res}\leq \gamma_{c1}$, we will adopt $<$ since the 
actual luminosity could be much lower than the predicted upper limits.

The complicated form in equation (\ref{Lxi}) could be simplified
for the pulsars in which pair production from ICS photons are not
important. This is the case for most pulsars except the ones with
high $B$ (Table 2). For $\zeta_{\rm ICS,k} < 2$, equation (\ref{Lxi})
turns out
\begin{equation} 
L_{x,nth}(E_c)\lesssim L_{\rm pc}\sum_{k=1}^{\rm int(\zeta_{\rm
SR})} (\eta_\perp^{k-1} \eta_\parallel \eta_{c,k,1}).
\label{Lxi2} 
\end{equation} 
In our calculation, we have adopted (\ref{Lxi}) uniformly without
losing generality, and used (\ref{Lxi2}) to do the comparison.
We found that, for high $B$ pulsars, the luminosity from (\ref{Lxi})
is much higher than the one from (\ref{Lxi2}), which means that the
higher order generations have important contributions to these pulsars.

The use of (\ref{Lxi}) for the millisecond pulsars is not accurate.
This is because, for millisecond pulsars, we have
$\gamma_c > \gamma_{1,1}$. In other words, the maximum emission energy
of the first ICS-branch (i.e. the ICS-branch from the secondary pairs)
has dropped below the ASCA and ROSAT band. This can be understood
by noting the very small $\eta_\parallel({\rm msPSR})$
(eq.[\ref{etamspsr}]) and $\kappa_{\rm ICS}({\rm msPSR})$
(eq.[\ref{kappaicsmspsr}]). Thus using (\ref{Lxi}) will
underestimate the non-thermal X-ray luminosities. To amend this,
we introduce a separate formula for millisecond pulsars
\begin{equation} 
L_{x,nth}^{\rm msPSR}(E_c)\lesssim L_{\rm pc}\sum_{k=1}^{\rm
int(\zeta_{\rm SR})} (\eta_\perp^{k-1}\eta_\parallel^{\zeta_{c,k-1}}),
\label{Lxni} 
\end{equation} 
where $\zeta_{c,k}$ is defined by
\begin{equation} 
\zeta_{c,k}=\left\{ 
		\begin{array}{ll} 0, & E_c\geq E_k \\ {\log(E_c/E_k)
		\over \log \kappa_{\rm ICS}}, & E_c < E_k \\
		\end{array} \right. ,
\label{zetac} 
\end{equation} 
$E_c$ is the energy we are interested in, and $E_k$ follows
(\ref{Ek}). $\zeta_{c,k}$ is the number of the ICS generations
with which the typical emission energy is dropped to $E_c$ from
$E_k$. For millisecond pulsars, this value is smaller than unity.
Thus equation (\ref{Lxni}) can raise the X-ray luminosity estimate
a little bit compared with (\ref{Lxi}) or (\ref{Lxi2}).
Note again the adoption of $<$, $\lesssim$ or $\sim$ is pending 
on the resonant condition.
 
Finally, we should notice another general relativistic effect, i.e.,
the redshift of the photons, when comparing the theoretical
predictions with the observations.  More specifically, the energy of
interest $E_c$, which is used in above calculations, is connected with
the observational energy $E_{\rm obs}$ by a redshift factor:
\begin{equation} 
E_c=(1-{r_g \over r_e})^{-1/2} E_{\rm obs} \simeq (1-0.4
r_{e,6}^{-1})^{-1/2} E_{\rm obs}.
\label{redshift} 
\end{equation} 
The redshift correction factor $(1-{r_g\over r_e})^{-1/2}$ is 1.13 
for $r_{e,6}=1.8$ and 1.29 for $r_{e,6}=1$.  By adopting 
$E_{\rm obs}=0.7-10$keV and 0.1-2.4keV, we can then calculate the 
(maximum) non-thermal X-ray luminosities for the ASCA and ROSAT 
bands, respectively.

\subsection{Results}

We have performed detailed calculations of the predicted non-thermal
and polar cap heating thermal X-ray luminosities of all the known 
X-ray pulsars detected by ROSAT or ASCA. In
Table 2, we listed the model parameters of these 35 pulsars, which
are separately grouped into normal and millisecond pulsars. 
Table 3 shows the calculation results of both the thermal and 
non-thermal X-ray luminosity predictions of the known X-ray 
pulsars and the comparison with the observational data (BT97 
and S98). The model prediction for the total ROSAT band is a sum 
of the ICS-origin non-thermal luminosity ([\ref{Lxi}] or [\ref{Lxni}]),
the thermal luminosity due to polar cap heating ([\ref{Lx-th-I}]
or [\ref{Lx-th-II}]), and the cooling thermal luminosity by assuming 
$R=10^6$cm for all the pulsars. This 
is because the predicted peak thermal components usually fall into 
the ROSAT band. For non-thermal luminosity both for ROSAT and ASCA
bands, we use (\ref{Lxi}) for all the normal pulsars, and use
(\ref{Lxni}) for all the millisecond pulsars.
Unless the temperature data are available from data analysis 
($T_{s,6}\sim 1.5$ for Vela, \"Ogelman et al. 1993; $T_{s,6}\sim 
0.56$ for Geminga, Halpern \& Wang 1997; $T_{s,6}\sim 0.8$ for both
PSR 0656+14 and PSR 1055-52, Greiveldinger et al. 1996, Wang
et al. 1998), we have adopted the values of $T_{s,6}$ and
$T_{h,6}$ according to (\ref{cooling}) and (\ref{T-I}) or (\ref{T-II}). 
For other model parameters, we have adopted $\xi=1$ (the last 
open field line), $\alpha=30^{\rm o}$ for all the pulsars, 
$R_{E,6}=1.8$ for most of the normal pulsars, $R_{E,6}=1.5$ for 
the relatively old normal pulsars (PSR 0950+08 and PSR 0823+26) and 
$R_{E,6}=1.2$ for the the millisecond pulsars. Such adoptions for 
$R_{E,6}$, though rough, are based on the numerical results of HM98
(their Fig.10). The weak dependence of $R_{E,6}$ on pulsar parameters 
like $P$, $B_p$ and $T$ is the main reason for us to adopt
fixed values of $R_{E,6}$ for different groups of pulsars.
In principle, for normal pulsars, $R_{E,6}$ is not a free parameter, 
since it is determined by $T_{h,6}$ and the 
pulsar parameters ($P$, $B_{p,12}$ and $\cos\alpha$), and $T_{h,6}$ 
itself is also the function of $P$, $B_{p,12}$ and $\cos\alpha$ 
(see eqs.[\ref{T-I}] or [\ref{T-II}]) if we assume the polar cap heating 
is the only process to decide polar cap temperature. However, the 
complication of the processes prevents an analytic expression of 

\centerline{}
\centerline{\psfig{figure=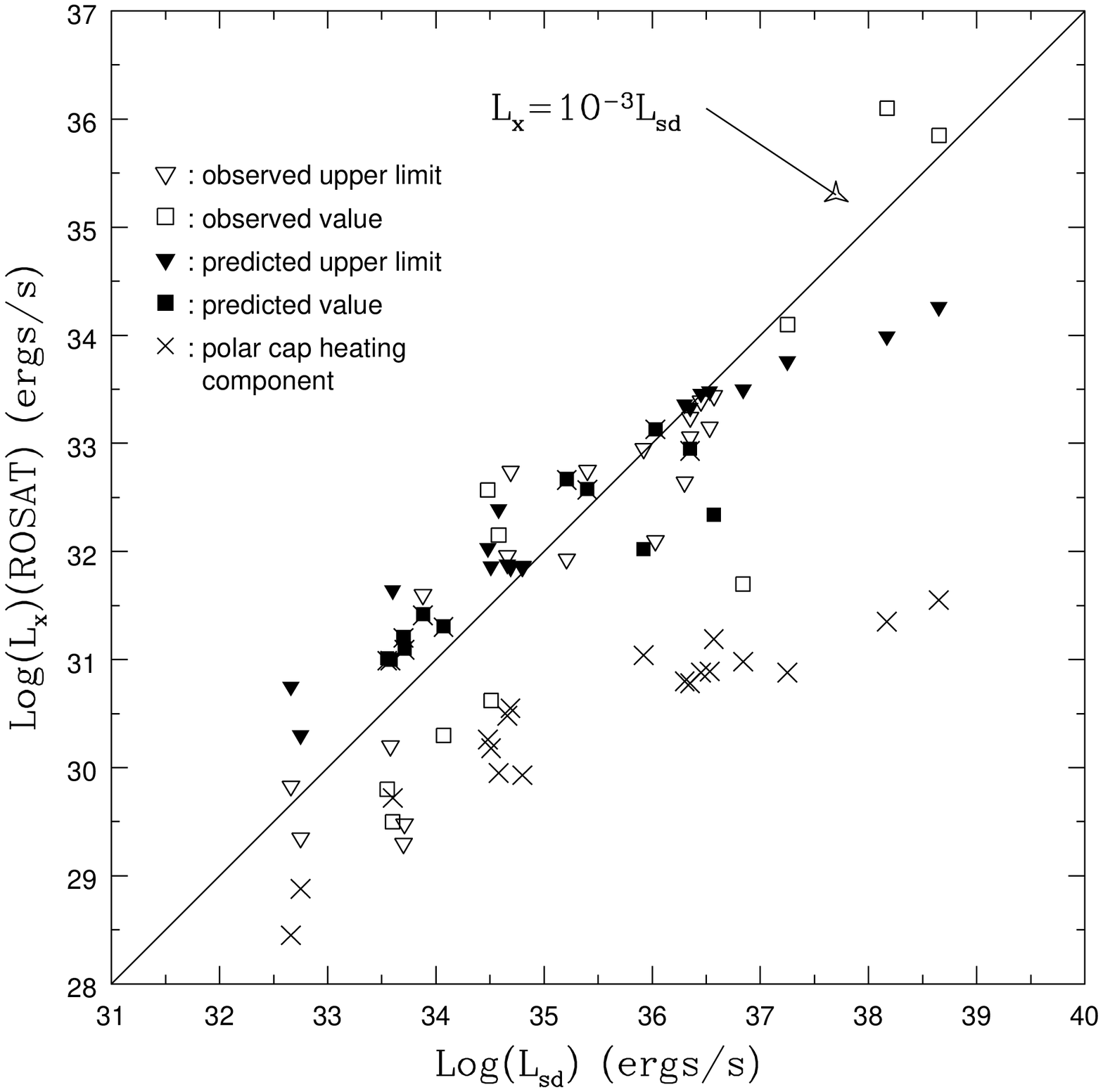,width=8.0cm}} 
\figcaption{Comparison of the luminosities observed and predicted in 
the ROSAT band for 29 pulsars. Note upper limits are adopted for some 
pulsars both for the observational and theoretical values. Note how
$L_x(ROSAT)\propto L_{\rm sd}$ feature is well reproduced, and how
the thermal component due to polar cap heating dominates the luminosity
in millisecond pulsars.
\label{fig:rosat}}
\centerline{}
     
The X-ray luminosity predictions versus observations are plotted in
Fig.4 and Fig.5.  If a pulsed signal is detected for a certain source,
we have adopted the {\em pulsed} emission luminosity as the observational
value, since we believe that only this luminosity is relevant to what we
are discussing in this paper. For the detections without pulsed signals,
we adopt their luminosities as upper limits. For the theoretical
results, we adopt the sum of both the thermal and non-thermal X-ray
luminosities for the ROSAT band, but only non-thermal luminosity for
the ASCA band. We regard both the values marked with `$\sim$' and
`$\lesssim$' in Table 2,3 as the actual values, while the values
marked with `$<$' as upper limits to plot Fig.4 and Fig.5. It is
notable from Fig.4 that, a $L_x(ROSAT)\propto L_{\rm sd}$ feature 
is roughly reproduced, and the thermal component due to polar cap 
heating clearly dominates the ROSAT band luminosities of the 
millisecond pulsars. The comparison of the theoretical predictions with 
the ASCA observations (S98, Saito et al. 1998) is less satisfactory, but
not in strong confliction when one bears in mind the small sample and
large uncertainties in the ASCA data. Furthermore, the inconsistency 
could be weakened by adjusting free parameters like $\alpha$ and 
$R_{E,6}$.

\centerline{}
\centerline{\psfig{file=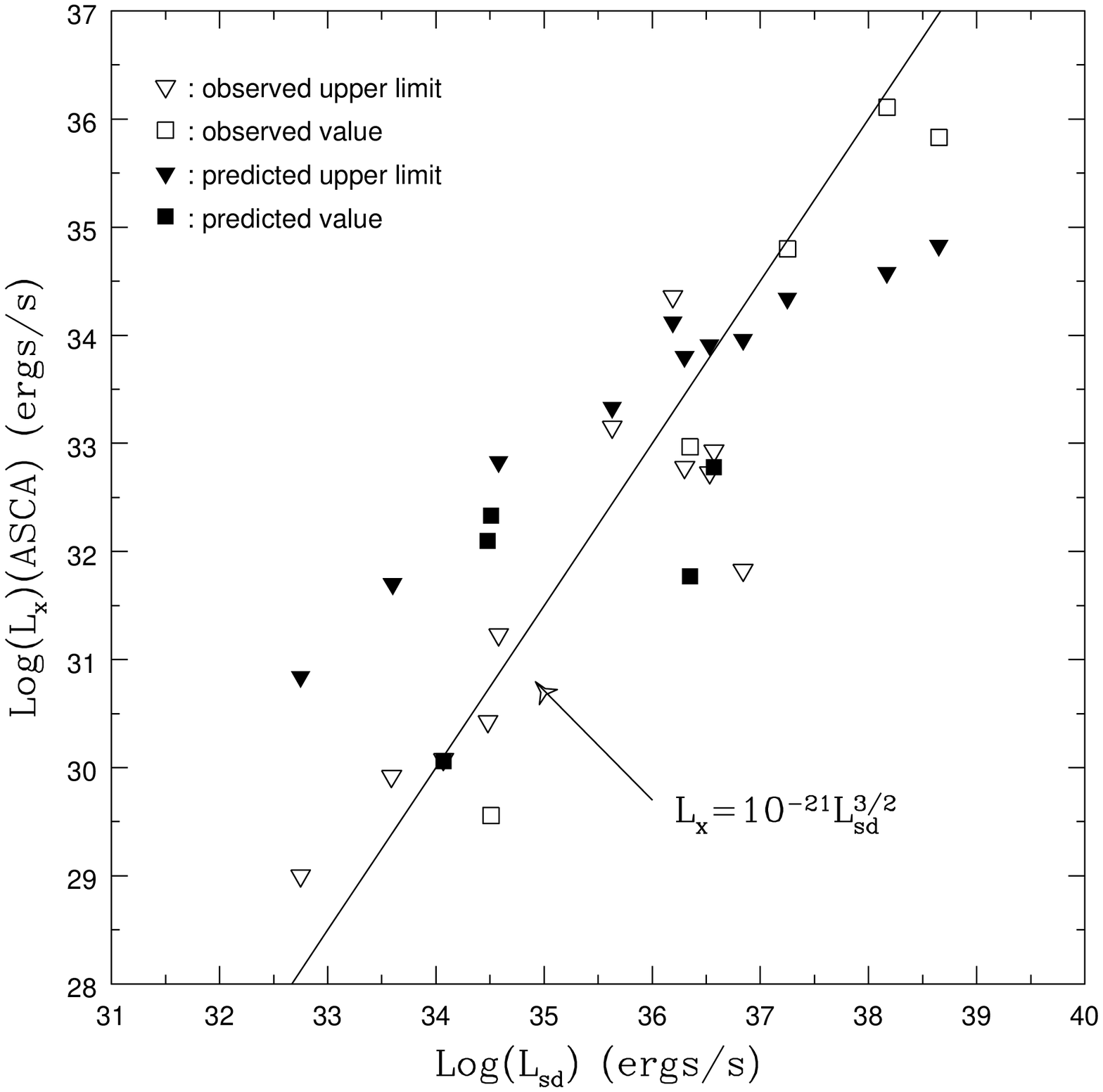,width=8.0cm}} 
\figcaption{Comparison of the luminosities observed and predicted in the 
ASCA band for 15 pulsars. Notations are same as those adopted in Fig.4.
\label{fig:asca}  }
\centerline{}
 
In Fig.6 and Fig.7, we plotted the dependence of our model predictions
on the free parameters $\alpha$ and $R_{E,6}$. We see both the
$\gamma$-ray and X-ray luminosities are mildly decreasing when
$\alpha$ gets larger. The decreasing becomes prominent when $\alpha$
gets closer to 90$^{\rm o}$. However, one should notice that such
decreasing can not get too low (say, below 2 orders of magnitudes),
since the smaller $\sin\alpha$ term, which has been neglected in
writing down the parallel electric fields in the HM98 model, will
become important for $\alpha\sim 90^{\rm o}$. The dependences of
the non-thermal X-ray luminosity on $R_{E,6}$ or $\alpha$ are not 
smooth (Fig.6a,7a,7b). This is expected since the zigzags actually
reflect the contributions from different ICS branches. For different
adoptions of $R_{E,6}$ of $\alpha$, the significance of the 
contributions of certain ICS branches to the final luminosity in a 
certain band actually varies. Thus the zigzag features are real 
physical effects. However, we notice that there exist several
non-differentiable points in the curves.
This is an artifact, which arises from the truncation
features in (\ref{etackj}). We believe that the detailed numerical
simulations can smooth out these non-differential points, though the
small ``bump'' features should remain. The present estimate
could bring some errors near these non-differential points. For the
millisecond pulsars, the height dependence is simpler (Fig.7b),
mostly because the generation structure is simpler (both
$\zeta_{\rm SR}$ and $\zeta_{\rm ICS}$ are smaller than 2), and
also due to the non-integer description (\ref{Lxni}).

\begin{figure*} 
\centerline{\psfig{file=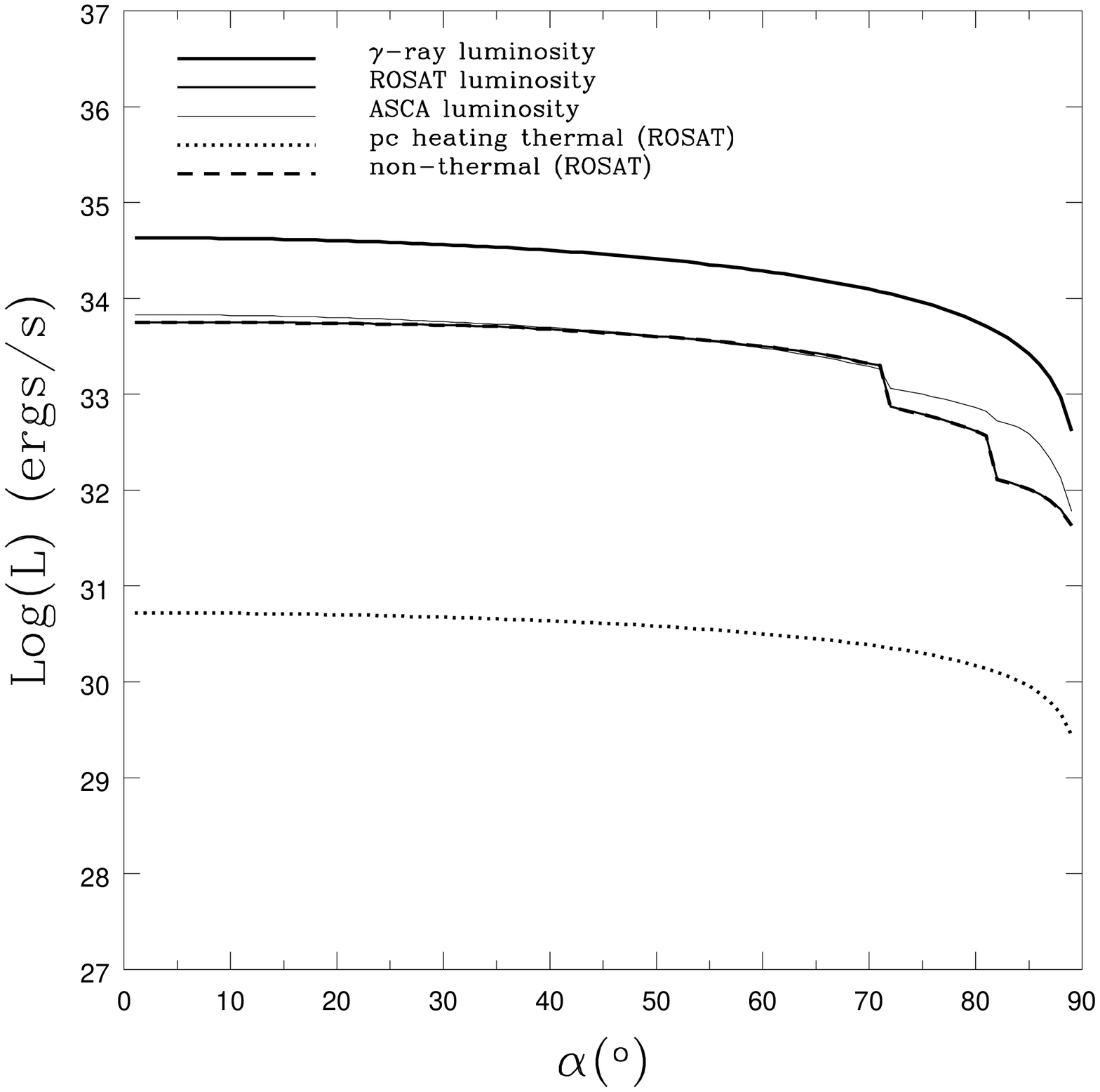,width=8.0cm}
 \psfig{file=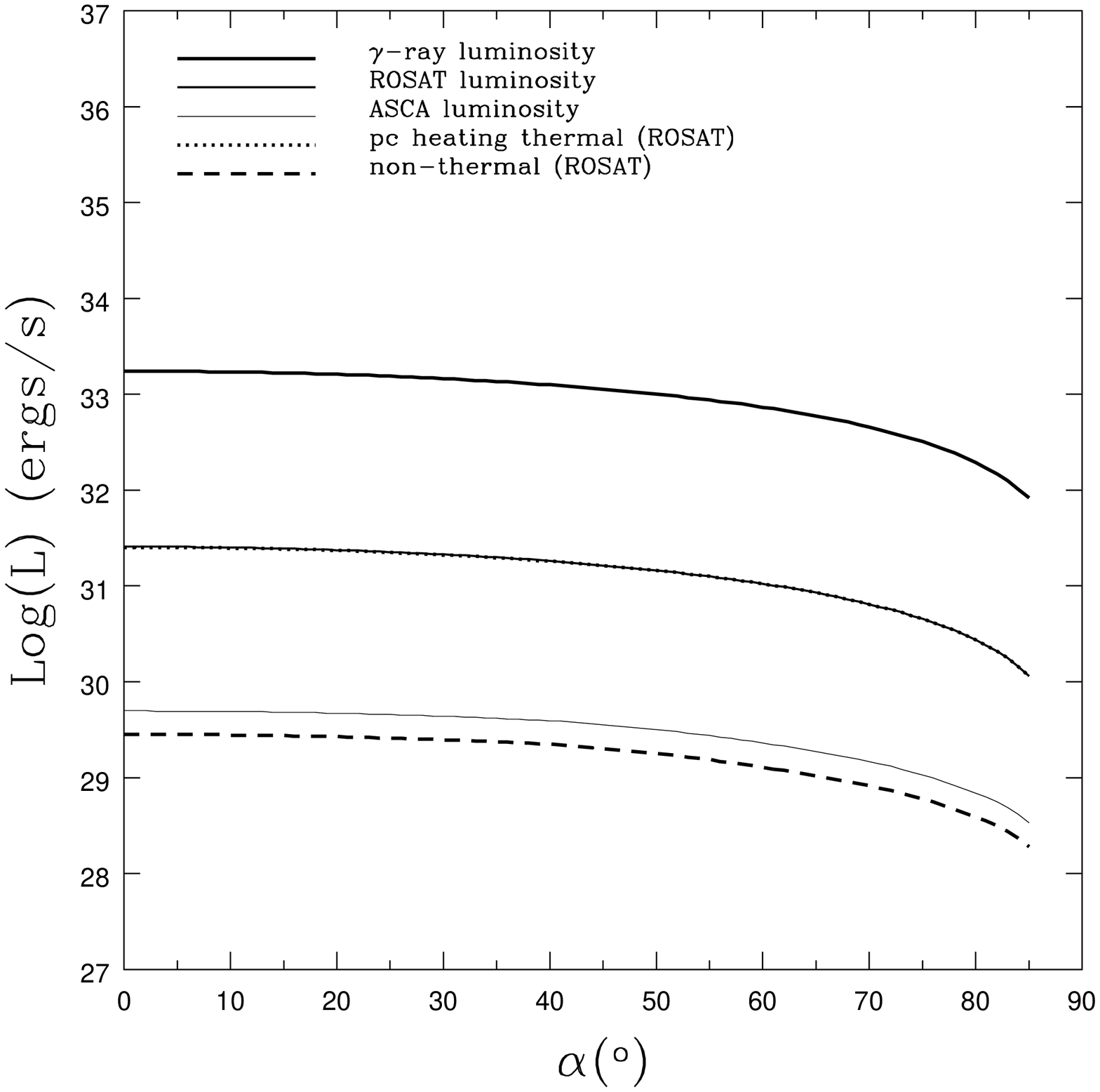,width=8.0cm}}
\caption{The inclination angle ($\alpha$) dependence of the predicted  
luminosities. (a) for a normal pulsar: $P=0.1$s, $B_{p,12}=5.8$,  
$r_{e,6}=1.8$, note ROSAT luminosity is dominated by the non-thermal  
component; (b) for a millisecond pulsar: $P=0.005$s, $B_{p,12}=5\times  
10^{-4}$, $R_{E,6}=1.2$, note ROSAT luminosity is dominated by the  
thermal component.
\label{fig:alpha} }
\centerline{}
\end{figure*} 
 
\begin{figure*} 
\centerline{\psfig{file=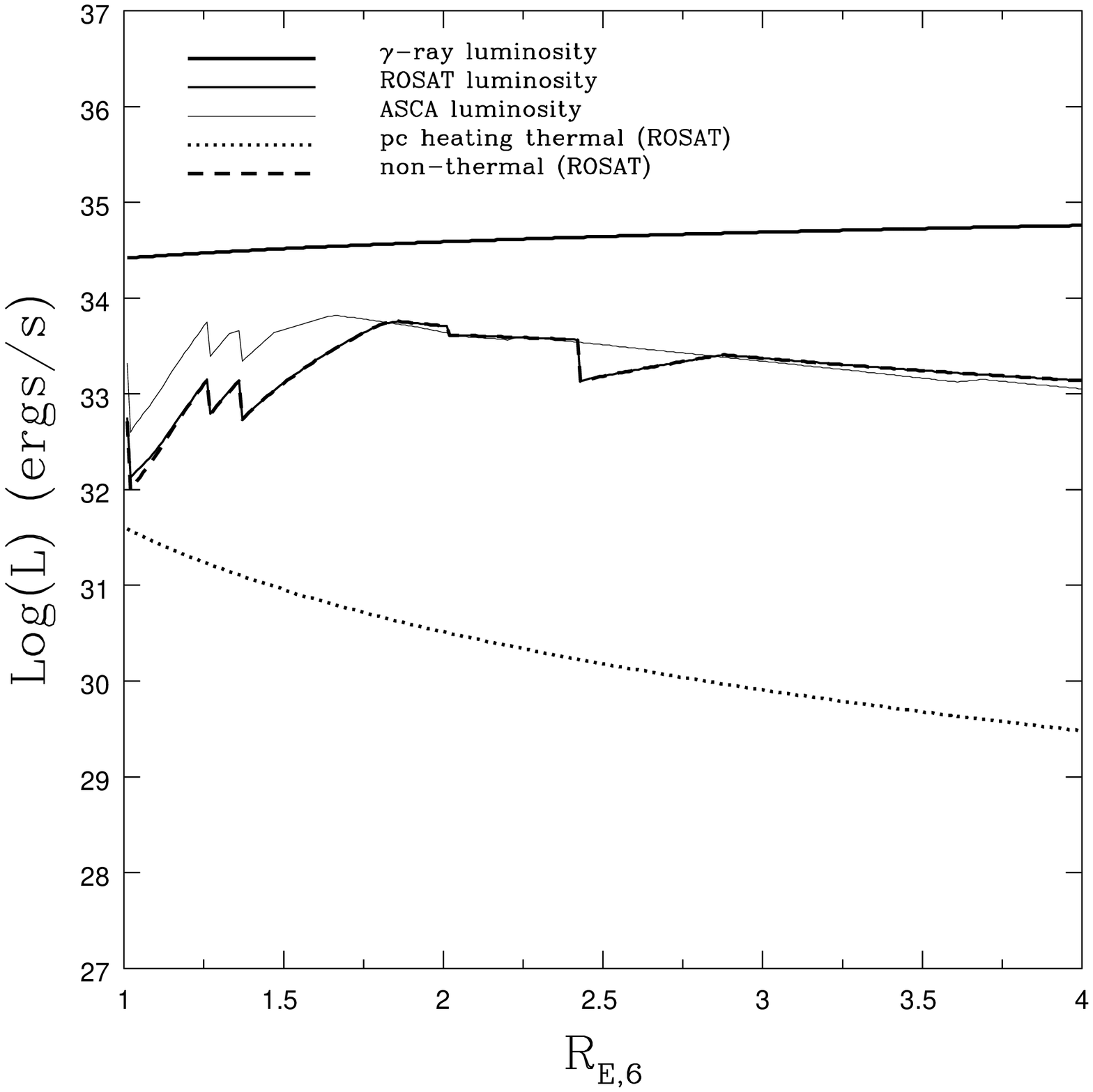,width=8.0cm}
  \psfig{file=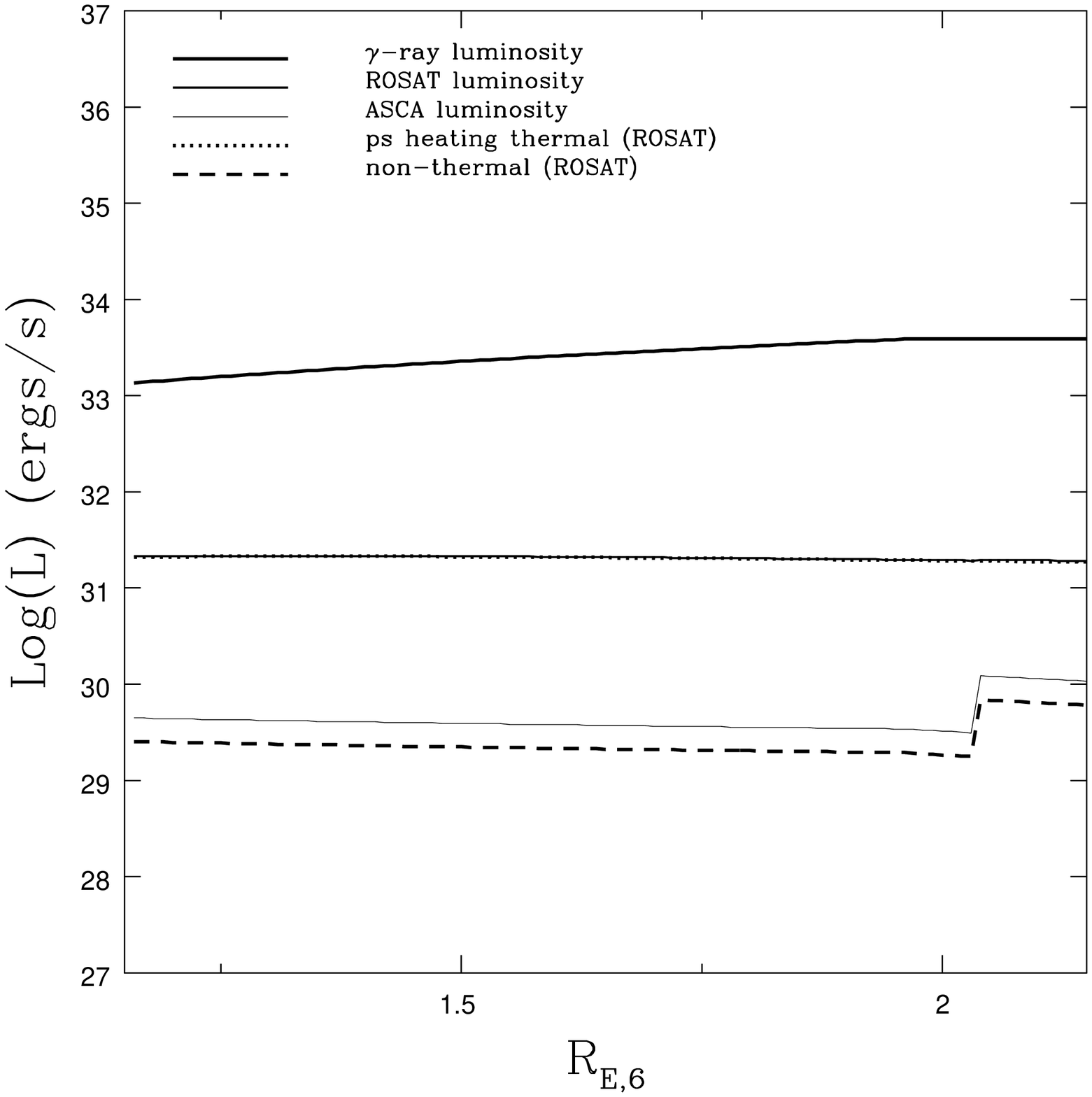,width=8.0cm}}
\caption{The emission height ($R_{E,6}$) dependence of the predicted  
luminosities. (a) For a normal pulsar: $P=0.1$s, $B_{p,12}=5.8$,  
$\alpha=30^{\rm o}$. b) For a millisecond pulsar: $P=0.005$s, 
$B_{p,12}=5\times 10^{-4}$, $\alpha=30^{\rm o}$.  The zigzags indicate
the contributions from different ICS branches. Note also that the ROSAT 
luminosity is dominated by the non-thermal component for normal pulsars, 
but by thermal component for millisecond pulsars.
\label{fig:re6}}
\centerline{}
\end{figure*}

\section{Conclusions and discussions} 

\subsection{Conclusions}

We have modified the conventional CR(or ICS)-SR polar cap
cascade picture by including the contributions of the ICS photons
produced by the higher generation pairs within the framework of the
space-charge-limited flow acceleration model proposed by Harding \&
Muslimov (HM98). Such a full-cascade picture is more complicated than
the canonical model. The complications mainly result from the
two-branch feature of the cascade, and the different regimes of the
ICS process under different conditions.  The important parameters,
i.e., the temperatures of both the full surface and the hot polar cap,
are still uncertain since there are many other factors (besides
cooling and polar cap heating) that can influence them. The
uncertainty of the temperatures will bring uncertainties of the
``resonant ICS condition'' (eq.[\ref{res}]), the height of the
accelerator $R_{E,6}$, etc., and thus bring uncertainties to our model
predictions. Furthermore, the complication of the accelerator itself
(HM98) has already been difficult to delineate
analytically. Therefore, to achieve a more accurate prediction, one
has to appeal to detailed Monte Carlo simulations.  However, some
interesting features, e.g. the resonant condition is satisfied for a
wide range of parameters; $R_{E,6}$ does not sensitively depend on 
pulsar parameters; the field strength variation is not important within
generations; only ICS branches have notable contributions to X-ray
luminosities; and so on, make an analytic description of the process
possible. We have tried to treat the process as accurately as possible
so as to keep as much information as we could. Such efforts include
the correction of the general relativistic effects (enhancement of the
field strength and photon energy redshift); the correction of the
non-zero pitch angle effect for higher SR generations; more accurate
and separate adoptions of the critical $\chi$ for describing the
cascade and escaping processes; separate treatments of the normal and
the millisecond pulsars, and so on. As shown in the previous sections,
our analytic description has presented an overview to the high energy
emission features of the pulsar population. Our main findings in this
paper can be summarized as follows:
 
1. When high energy photons are converted to pairs in the pulsar
magnetosphere via $\gamma-B$ process, not only the perpendicular
energy of the pairs are further converted into radiation via SR, most
of the parallel energy of the pairs will be also converted to
radiation via the ICS of the particles with the soft thermal
photons. Thus there is almost no energy loss during the conversion of
the accelerating particle energy to radiation in the cascade process. 
Such a full-cascade scenario gives a natural interpretation to the 
$L_{\gamma}\propto (L_{\rm sd})^{1/2}$ feature.
 
2. SR branches of the cascade have little contribution to the X-rays 
in the ROSAT and ASCA bands. The ICS-branch spectra, however, can
result in a non-thermal X-ray component with their soft tails. 
Such non-thermal X-rays of ICS-origin play an important role to
account for the ROSAT and ASCA luminosities.
 
3. The polar cap heating from backflowing particles of the inner 
accelerator can contribute a hot thermal component with small area 
(e.g. polar cap), which also has a contribution to the ROSAT 
luminosity.
 
4. The combination of the thermal and the non-thermal components in
the model reproduces the reported $L_x(ROSAT)\sim 10^{-3} L_{\rm sd}$ 
feature (BT97). The non-thermal components predicted in the ASCA band 
is not in severe contradiction to the ASCA data (S98).
 
5. For the ROSAT band, the non-thermal components dominate the
luminosities for all the normal pulsars, while the upper limits of the
thermal luminosities are usually higher than the non-thermal 
luminosities for the millisecond pulsars. We thus predict
that the unidentified spectral features of the {\em pulsed} 
emission of the millisecond pulsars might be of thermal origin.

Zhang \& Harding (1999) have also presented similar conclusions. 
The main improvement of this paper is that we have included both the
unsaturated and the saturated cases of the space-charge-limited
flow accelerators.

\subsection{Comparisons with the outer gap model} 

The $L_x\sim 10^{-3}L_{\rm sd}$ feature has been also interpreted
in terms of the thick outer gap model by Cheng et al. (CGZ98; CZ99),
thus it is of interest to compare our model with the outer gap model.
The main differences between the two models lie in the following 
aspects:
 
1. We interpret the non-thermal X-rays as being due to the ICS of
upward polar cap cascades, while the outer gap model attributes the
non-thermal X-rays to the SR of the downward cascades from the outer 
gap particles. A prominant feature of our model is that we can 
simultaneously reproduce both the observed $L_{\gamma}\propto 
(L_{\rm sd})^{1/2}$ and the $L_x \sim 10^{-3}L_{\rm sd}$ dependences 
well. The thick outer gap model, however, has a quite different 
$L_{\rm sd}$-dependence of the $\gamma$-ray luminosities ($L_\gamma
\propto B_{p,12}^{2/7}P^{-2/7}$, see a combination of eqs.[23], [22] 
of Zhang \& Cheng 1997) which severely violates the observed 
$L_{\gamma}\propto (L_{\rm sd})^{1/2}$ dependece, although
$L_x \sim 10^{-3}L_{\rm sd}$ feature was also well reproduced.

2. As for the thermal X-rays, our model predicts a hot small area 
component in terms of polar cap heating, and adopt the full-surface
temperature from a simple cooling model. In the outer gap model, 
both the full-surface and the polar cap temperatures are predicted 
in terms of the thick outer gap heating (Zhang \& Cheng 1997; CGZ98; 
CZ99). CZ99 also took into account the possible effect of the polar 
cap heating in old pulsars PSR 1929+10 and PSR 0950+08, in which the 
outer gap might not be formed. An important difference is that, for
our model, as long as the resonant condition (\ref{res}) is satisfied, 
our model predictions are not sensitively dependent on the surface 
temperatures, while in the outer gap model, the surface temperatures
are closely related to the non-thermal X-ray luminosities due to
the self-consistency of the model. 
 
3. In the pure non-thermal-origin outer gap model (CGZ98), some direct
analytic expressions of the soft X-ray luminosities in terms of the
spin-down energy (their eqs.[15,16]) are available. However, such a
model is incomplete since the thermal emission should have a
non-negligible contribution to the ROSAT band emission, and such
thermal components have indeed been identified from some pulsars
(see Sect.1). CZ99 improved the model by incorporating the thermal 
contributions.  But they assumed that the hard polar cap thermal 
components do not have a significant contribution to the ROSAT
observations. Their results combining both the non-thermal and the
soft full-surface thermal components can also reproduce the 
$L_x(ROSAT)\sim 10^{-3}L_{\rm sd}$ feature. In our polar cap 
model, a simple expression for the hot thermal luminosity 
([\ref{Lx-th-I}] or [\ref{Lx-th-II}]) is available, but no simple 
analytic expression 
(in terms of pulsar parameters) could be achieved to describe the 
non-thermal luminosity due to complexity of the cascade processes.
A sum of both the thermal and non-thermal luminosities can also
results in a rough $L_x(ROSAT)\sim 10^{-3}L_{\rm sd}$ dependence. 
The distinguishing point between the two models might be the 
different origin of the soft X-rays from the millisecond pulsars. 
As addressed above, our model predicts that the pulsed emission of 
the ROSAT-observed millisecond pulsars might be dominated by the 
thermal component due to polar cap heating, while the outer gap
model attributes the millisecond pulsar soft X-rays also to the 
non-thermal SR component, although they adopted an additional 
assumption of strong multipole magnetic field components for these 
pulsars. Such an issue could be a discriminator between the two 
models. Unfortunately, among the 10 X-ray millisecond pulsars, 
spectral analyses are only performed on two pulsars. Although 
there is evidence of the existence of a hot thermal component 
in PSR J0437-4715 (Becker \& Tr\"umper 1993; Halpern, Martin \& 
Marshall 1996; Zavlin \& Pavlov 1998), a combination of the ROSAT 
and ASCA data shows that the spectrum of PSR B1821-24 is a power-law, 
and thus, of non-thermal origin.  According to our results (Table 3), 
the observational values both in ROSAT and ASCA band for PSR 
1821-24 are much higher than model predictions, and also much higher 
than the observed values of other millisecond pulsars. This leads us 
to suspect that there might be some special non-thermal X-ray 
emission mechanisms operating in this pulsar. Thus, when more data 
are accumulated, the spectral analyses of the pulsed emission of 
the other millisecond pulsars could provide evidence for or against 
our idea. It is worth noticing that the thermal feature for the 
millisecond pulsars is also expected by Zavlin \& Pavlov 1998). 
However, if there indeed exist strong multipole magnetic fields
near a millisecond pulsar's surface, the differentiating between
the models becomes difficult, since our model also predicts much
stronger non-thermal emission components due to the enhancement of
the ICS emission efficiency.

4. Among 35 detected spin-powered X-ray pulsars, pulsed emission
components are only observed in 10 pulsars by ROSAT, 4 of which
(Crab, Geminga, PSR B1509-58 and PSR B0540-69) are also observed by
ASCA. ASCA also observed pulsed strong non-thermal emission from
another pulsar, the millisecond pulsar PSR B1821-24. For other
pulsars, only the total X-ray luminosities are reported. Our polar
cap model actually predicts pulsed emission luminosities.
Thus, as long as the pulsed luminosity is detected, we use
it to compare with the model prediction. For those pulsars in
which no pulsed emission is detected, we regard the total luminosity
as an upper limit (see Fig.4,5), since besides the internal (cooling
and heating) and magnetospheric (polar cap or outer gap heating, and
ICS or SR non-thermal emission) origin X-rays, there are still many
mechanisms, such as the accretion from the interstellar medium
(Paczynski 1990), pulsar wind nebular emission (Arons \& Tavani 1993),
etc., which can account for non-pulsed X-ray emission. We noticed that,
however, in the outer gap model, the total luminosities are adopted to
compare with the model predictions. On the theoretical aspect, the
outer gap model predicts a precise value, but our model predictions 
could either be a actual value or an upper limit pending on whether 
the ICS occurs in the resonant regime (see Table 3; Fig.4,5). More
precise predictions will be available after detailed numerical 
simulations.

5. Besides the spectral behavior of the millisecond pulsars, 
another issue could also be regarded as a differential criterion 
of the competing models. For the relatively old canonical pulsars such 
as PSR 0950+08 and PSR 1929+10, the outer gap model predicts pure 
thermal emission (due to polar cap heating), since outer gaps could 
not be formed in these pulsars (CZ99). Our model, however, predicts 
a non-thermal plus thermal feature. The thermal emission component 
due to polar cap heating has a luminosity comparable to the 
ICS-origin non-thermal component, so that both of them are 
detectable. 

\subsection{Further remarks}

We have some further remarks on our model. Firstly, the analytic
description approach adopted in this paper has some limitations
in presenting detailed predictions. As shown in Sect.\ref{gop}, when
deducing the recursion relations between adjacent generations, we
have only examined the characteristic emission energies, but ignored
the detailed radiation spectrum of each generation. This is the main
reason for the non-smooth features in Fig.7a. Furthermore, the 
expressions of the predicted luminosities may need further revisions 
to include more effects that might be important. For example, in 
(\ref{Lxi}), one may need to incorporate a weighting factor to
describe the $\gamma$-dependence of $\dot\gamma$, since (\ref{Lxi})
assumes equal number density for the particles with different
$\gamma$s, which might not be necessarily the case. Therefore 
to get detailed $\gamma$-ray and X-ray spectra, careful numerical 
simulations are desirable. Nevertheless, we can give a rough 
estimate of the non-thermal X-ray spectral indices. According to
Dermer (1990), the thick-target magnetic ICS spectral indices are
approximately $1+\Gamma/2$ or $\Gamma$ within different energy regimes
(see his eqs.[41a,41b]), where $\Gamma$ is the spectral index of the
stable input particle spectrum. Monte Carlo simulation results (DH96)
show that for the higher generation pairs, usually one has $\Gamma 
\sim 1.5-2$, so that the final photon spectral indices are around 
$1.75-2$. This is in good agreement with the observations (BT97).

Our present model is not good at describing the high magnetic field
pulsars. Although we have tried to accomendate high $B$ pulsars by 
including higher order ICS branches (eq.[\ref{Lxi}]), there are 
still several aspects of high $B$ pulsars that we did not take 
into account. First, the model predictions presented in this pulsar 
are mainly very loose upper limits rather than actual values.
This is because, for high $B$ pulsars, the cyclotron energy
is very high. The typical Lorentz factor for the resonant ICS
condition (eq.[\ref{res}]) is thus also very high (e.g. $\sim 10^3$),
and is much higher than the Lorentz factor (e.g. about 100) required
to emit soft X-rays in the ROSAT or ASCA bands. Thus the electrons 
may not be able to reduce their energies enough to scatter the soft
photons to these bands, so that in some cases, the pulsar may not 
be detected in X-rays at all. Another point is that in high $B$ 
pulsars, the accelerator should be controlled by ICS, since both 
the upward and downward scatterings occur in the resonant regime,
and the anisotropy of the ICS processes no longer exist. The 
accelerator is hence moved back to the vicinity of the surface
($R_{E,6}=1$), and the expression of $E_0$ ([\ref{E0-I}] or 
[\ref{E0-II}]) should be modified by the ICS-controlled one.
All these will alter the generation structures of the high $B$ 
pulsars, and change the model predictions to some extent. 
Furthermore, photon splitting is also important for high $B$ pulsars 
(HBG97), and will complicate the cascade picture even more. Thus to 
compute the cascade spectrum of these high $B$ pulsars, one has to 
appeal to rather complicated numerical simulations (see e.g. 
Baring \& Harding 1999).

Though with limitations, the analytic approach adopted in this paper
has its advantage to present a general overview of the emission 
properties of different pulsars without performing detailed numerical 
simulations. For example, our model shows that, for most young pulsars,
the non-thermal luminosities are much higher than both the full-surface
and the hot polar cap thermal luminosities, so that the thermal
components should be buried by the non-thermal power law continuum.
For some middle-aged pulsars such as Geminga, PSR 0656+14, and PSR 
1055-52, however, the predicted non-thermal X-ray luminosities are 
of the same order of the full-surface thermal luminosities, thus 
these thermal components could be detectable. This is just what is 
observed. Another example is that our model seems 
to have the ability to interpret some ``missing detections''. For 
example, ROSAT has detected 10 millisecond pulsars. If pulsar X-rays 
for the millisecond pulsars are indeed of non-thermal origin as the 
outer gap model argued, then most of them should be detected by ASCA 
as well, since the ASCA's band is higher and wider. However, only 2 
out of those 10 are detected by ASCA. In our model, the ROSAT-band 
emission for these millisecond pulsars is dominated by the thermal 
emission, which is stronger than the non-thermal component. In the 
ASCA band, only the non-thermal component is present, so that the 
predicted luminosities are lower. In fact, our predictions of the 
ASCA luminosities of these 10 millisecond pulsars are usually too 
low to be detected (see Table 3) except for PSR B1821-24, which has 
been indeed observed to be dominated by the non-thermal emission 
though our prediction is still less by almost an order of magnitude. 
For PSR J0437-4715, we predicted a low luminosity in ASCA band, 
which is close to the observational level. The detection of this 
pulsar by ASCA is due to the close distance of the source. Another 
example of the ``missing detections'' is, among the ASCA detected 
16 pulsars (S98), only two pulsars, i.e., PSR B1610-50 and 
PSR 1853+01, were not detected by ROSAT. This is not easy to be 
understood in terms of the outer gap model. In our model, we can 
present a simple interpretation. We noticed that these two pulsars 
all have the magnetic fields higher than $10^{13}$G. The absence of 
soft X-rays from these two pulsars might just be due to the failure 
of the resonant scattering condition in the ROSAT band as discussed 
above. However, some other high $B$ pulsars, e.g. PSR B1509-58 and 
PSR B2334+61, have been detected by ROSAT. Thus numerical simulations 
are desirable to tell the differences in these pulsars.

There is growing evidence that the optical emission of the 
spin-powered pulsars are of non-thermal origin. The ICS of the
higher generation pairs discussed in this paper has a spectrum
which could extend to as low as the optical band. However, the
resonant ICS condition fails in this regime, so that we could 
not present a plausible estimate of the optical luminosities 
without detailed numerical simulations. Thus whether or not the 
observed non-thermal optical emission is due to such ICS processes 
remains unclear.

Radio emission from pulsars is commonly believed to be of certain
coherent origin owing to the very high brightness temperatures
observed. Thus pair production, which can arise various
plasma instabilities in the magnetospheres, is believed to be the
essential condition of pulsar radio emission. Our full-polar cap
cascade model presents multi-component pairs with different energies,
which increases the possibilities of the ``overtaking'' two-stream
instabilities that might account for pulsar radio emission.
Such detailed models deserve constructing.

\subsection{Detailed comparisons with observations}
 
It is worth presenting a closer comparison of our model
predictions with the observations of some well-observed pulsars.
 
Crab and PSR B0540-69: Crab and its twin in the LMC are the youngest
known pulsars.  Both the ROSAT and ASCA observations show that they
have very high luminosity radiation in X-rays, which are 1-2 orders
of magnitude higher than our predictions. This leads one to suspect
that there might be some more efficient X-ray emission mechanisms
operating in these pulsars. However, our model could not be ruled out.
We noticed that the luminosity data reported by BT97 and S98 are the
values assuming isotropic emission of the X-rays. Thus the actual
values should be much smaller if the radiation is strongly beamed.
For example, these values could be down by an order of magnitude if
one assumes the radiation is coming from 1 steradian solid angle. 
For Crab and PSR B0540-69, the strong non-thermal spectral features 
make us suspect its magnetospheric origin, and thus, the emission 
could be beamed.

Vela: Vela pulsar is actually weaker in X-rays than expected.
It was not until 1993 when \"Ogelman, Finley \& Zimmermann (1993)
first discovered its weak pulsed X-ray emission with ROSAT. Its soft
X-ray emission is dominated by a thermal component with the surface
radius only 3-4km, which is much smaller than the conventional value,
i.e. 10km. In fact, in BT97 and S98, the detected X-ray luminosities
are much lower than the expected values according to their empirical
relationships.  Recent RXTE (2-30keV) observations (Strickman, Harding
\& deJager 1999) reveal an even lower luminosity in this band.
Such a feature does not conflict with our model prediction, since 
what we predicted is an upper limit (see Table 3).
An interesting discovery of Strickman, Harding \& deJager (1999) is
that although the RXTE spectrum of Peak 1 of this pulsar joins
to the OSSE spectrum smoothly, the spectrum of Peak 2 does show
the evidence of another component, which would extrapolate to within
a factor of two of the optical spectrum. Detailed numerical
simulations might be able to show whether the extra component is
of ICS origin.

PSR B1509-58 and other high $B$ pulsars: We have discussed the
possible complications in high $B$ pulsars. Our present predictions
are not inconsistent with the observations. 
It is worth noticing that by adopting (\ref{Lxi2}) rather
than (\ref{Lxi}) to do the calculation (thus assuming ICS branches
do not pair produce), the prediction is nearly 2 order of magnitudes
lower than the observations for PSR B1509-58. This indicates
that the higher order ICS branches actually play an important
role in high $B$ pulsars.

Geminga, PSR B1055-52, and PSR B0656+14: these three middle-aged
pulsars are the ones which have detailed spectral information, and 
are identified as ``cooling pulsars'', which consist of a cooling
thermal component and a non-thermal power law. As shown in Table 2,
our model predictions show comparable luminosities of the
non-thermal and full surface thermal components for these pulsars,
and hence can well reproduce the observed features. 

PSR 1929+10 and PSR 0950+08: As discussed in Sect.1, the spectral
nature of these two old pulsars are still unclear, though they
all probably have a hot polar cap thermal component. Since both the
full surface temperature and the hot polar cap temperature are much
lower in these pulsars, the ICS conversion efficiency should be
considerably lower. Thus the non-thermal luminosity could be much
lower than the upper limit we predicted. As a result, most probably,
the thermal component from the hot polar cap may have comparable
luminosity as the non-thermal component, so that such hot polar caps
could be detectable. Though PSR 1929+10 is fitted
by a single small area thermal component by Wang \& Halpern (1997), it
could be equally well fitted by a thermal polar cap plus a non-thermal
component spectrum (BT97), which is just expected by our model. Future
XMM observations on these two old pulsars will test our model 
predictions. As discussed in Sect.4.2, the thermal plus non-thermal
feature of these two pulsars might be also adopted as a distinguishing
criterion for the competing models.

Millisecond pulsars: as discussed above, our model prediction for 
the millisecond pulsars could be a feature to distinguish the 
rivaling models. For PSR J0437-4715 and PSR B1821-24, our model 
seems to have presented a qualitatively correct picture.
The fact that a two component fit for PSR J0437-4715
requires a small-area thermal component actually supports our idea
of the thermal origin of the pulsed soft X-rays from millisecond 
pulsars. We emphasize that the thermal feature we have predicted is
only for the {\em pulsed} X-ray emission from millisecond pulsars.
Another remark is that stronger multipole fields near the surface,
if any, will make the polar cap model and the outer gap model 
indistinguishable since both models predict the non-thermal dominant 
feature.
 
We are grateful to the referee Werner Becker for his informative
instructions and helpful advice, and Alex Muslimov for important
discussions and useful comments. 
 
\clearpage
\appendix{ 
\section{Derivation of equation (35)} 
The point of this appendix was first noted by Cheng \& Ruderman
(1977), and was adopted in WSL97. However, as shown below, a slightly
different definition (i.e. A3, A4) makes (A9) an accurate relation
rather than an asymptotic form for the relativistic cases.
 
Suppose the $i$th generation $\gamma$-ray $\epsilon_{ i}$ produce
$(i+1)$-th generation pair with $\gamma_{i+1}=\epsilon_{ i}/2$ and
pitch angle $\theta_{\rm kB}$, then $\gamma_{i+1}$ can be
generally expressed as
\begin{equation} 
\gamma_{i+1}=[1+(p_{i+1,\parallel}/mc)^2+2nB']^{1/2},  
\end{equation} 
where $n$ is the number of the Landau levels in the direction
perpendicular to the field line. For $n\gg 1$, this could also be
expressed classically as
\begin{equation} 
\gamma_{i+1}=[1+(p_{i+1,\parallel}/mc)^2+(p_{i+1,\perp}/mc)^2]^{1/2},  
\end{equation} 
where $p_{i+1,\parallel}=p_{i+1}\sin\theta_{\rm kB}$ and
$p_{i+1,\perp}=p_{ i+1}
\cos\theta_{\rm kB}$. Define 
\begin{equation} 
\hat{\gamma}_{ i+1,\parallel}=[1+(p_{ i+1,\parallel}/mc)^2]^{1/2} 
\end{equation} 
and
\begin{equation} 
\gamma_{ i+1,\perp}=[1+(p_{ i+1,\perp}/mc)^2]^{1/2},  
\end{equation} 
we get
\begin{equation} 
\gamma_{i+1}^2=\hat{\gamma}_{i+1,\parallel}^2+\gamma_{i+1,\perp}^2-1  
\end{equation} 
and
\begin{equation} 
\cos^2\theta_{\rm kB}={\hat{\gamma}_{ i+1,\parallel}^2-1 \over 
\gamma_{ i+1}^2-1}. 
\end{equation} 
This is a little bit different from Cheng \& Ruderman (1977) who
defined $\hat{\gamma}_{ i+1,\parallel}=p_{ i+1,\parallel}/mc$ and
$\gamma_{ i+1,\perp}= p_{ i+1,\perp}/mc$. Note $\hat{\gamma}_{
i+1,\parallel}$ is not the parallel energy of the particle we are
interested in, since when the particle emits the SR photons, it will
receive a recoil force so that its momentum is usually changed.
 
Now define a Lorentz frame with $\gamma_{ i+1,\parallel}$ (different
from $\hat{\gamma}_{ i+1,\parallel}$) in which only perpendicular
momentum is left.  In such a co-moving frame, the emission of SR
photons will usually bring no recoil to the particle in the parallel
direction as long as the particle emits photons in the direction
perpendicular to the magnetic field, i.e., the particle is in high
enough Laudau levels. This is true in most cases we discussed in this
paper. (Note when the particle is in a very low Landau state, above
conclusion is not necessarily true, since the SR photons will be
emitted in random directions.) Thus such a Lorentz factor is identical
to the real Lorentz factor of the particle after it emits all its
perpendicular energy via SR. A handy Lorentz transformation leads
to the relation
\begin{equation} 
\cos^2\theta_{\rm kB}={1-\gamma_{ i+1,\parallel}^{-2} \over 
1-\gamma_{i+1}^{-2}}. 
\end{equation} 
Compare [A6] with [A7], we finally get
\begin{equation} 
\gamma_{i+1,\parallel}^2={\gamma_{ i+1}^2 \over \gamma_{ i+1}^2 
-\hat{\gamma}_{ i+1,\parallel}^2+1}
\end{equation} 
or
\begin{equation} 
\gamma_{ i+1,\parallel}={\gamma_{ i+1}\over\gamma_{ i+1,\perp}}=
{\gamma_{i+1} \over[1+(\gamma_{i+1}^2-1)\sin^2\theta_{\rm kB}]^{1/2}}, 
\end{equation} 
which is the equation (35) in the text.

\clearpage
  
\begin{table} 
\caption{Broad-band high energy luminosities (above 1eV, most 
luminosity contributions are from the band above 100KeV, thus 
denoted as $L_\gamma$) observed (Thompson et al. 1999) and expected. 
The luminosity for PSR B1046-58 is adopted by that above 400 MeV, 
following Kaspi et al. (1999). Pulsar data are taken from Taylor 
et al. (1995). `o' denotes observed value, `m' denotes model 
prediction, `f' denotes the full-cascade model discussed in 
this paper, and `c' denotes the canonical CR-SR cascade model.}
\begin{tabular}{lrccccc} 
\hline 
\hline 
Pulsars & P(ms) & Log B(G) & $L_{\rm sd}$ & $L_\gamma$(o) &
$L_\gamma$(m,f) & $L_\gamma$(m,c)\\
\hline 
$B0531+21$ & 33.40 & 12.88 & 38.65 & 35.70 & 35.54 & 34.94 \\
$B0833-45$ & 89.29 & 12.83 & 36.84 & 34.38 & 34.71 & 34.22 \\
$B0633+17$ & 237.09 & 12.51 & 34.51 & 32.98 & 33.56 & 33.38 \\
$B1706-44$ & 102.45 & 12.79 & 36.53 & 34.84 & 34.57 & 34.12 \\
$B1509-58$ & 150.23 & 13.49 & 37.25 & 35.21 & 34.85 & 32.33 \\
$B1951+32$ & 39.53 & 11.99 & 36.57 & 34.40 & 34.64 & 34.58 \\
$B1055-52$ & 197.10 & 12.34 & 34.48 & 33.79 & 33.56 & 33.45 \\
$B1046-58$ & 123.65 & 12.84 & 36.30 & 34.34 & 34.46 & 33.96 \\
\hline 
\end{tabular} 
 
\end{table}

\begin{table} 
\caption{Model parameters of the spin-powered X-ray pulsars. Pulsar 
 data are taken from Taylor et al. (1995). The 4th column `regime'
 denotes the regime of the accelerator: `I' indicates the unsaturated
 case, while `II' indicates the saturated case. Normal pulsars and 
 millisecond pulsars are grouped separately.}
 
\begin{tabular}{lrcclclccll} 
\hline 
\hline 
Pulsars & $P$(ms) & $\log B$(G) & ${\rm regime}$ & $f$ &
$\kappa_{\rm SR}$ & $\kappa_{\rm ICS}$ &
$\zeta_{\rm SR}$ & $\zeta_{\rm ICS}$ & $\eta_\perp$ & $\eta_\parallel$
\\
\hline 
B$0531+21$ & 33.40 & 12.88 & I & 0.00010 & $6.25\times10^{-2}$ 
& $9.66\times10^{-3}$ & 3.44 & 2.46 & 0.669 & 0.331 \\ 
B$0833-45$ & 89.29 & 12.83 & I & 0.00019 & $6.25\times10^{-2}$ 
& $7.64\times 10^{-3}$ & 3.21 & 2.26 & 0.704 & 0.296 \\ 
J$1811-1926$ & 64.66 & 13.11 & I & 0.00011 & $6.25\times10^{-2}$ 
& $2.58\times 10^{-2}$ & 3.41 & 2.83 & 0.485 & 0.515 \\
J$1617-5055$ & 69.33 & 12.80 & I & 0.00017 & $6.25\times10^{-2}$ 
& $6.71\times10^{-3}$ & 3.25 & 2.25 & 0.722 & 0.278 \\
B$0633+17$ & 237.09 & 12.51 & II & 0.00042 & $6.25\times10^{-2}$ 
& $1.74\times 10^{-3}$ & 2.62 & 1.71 & 0.856 & 0.144 \\  
B$1706-44$ & 102.45 & 12.79 & I & 0.00021 & $6.25\times10^{-2}$ 
& $6.56\times 10^{-3}$ & 3.16 & 2.19 & 0.725 & 0.275 \\ 
B$1509-58$ & 150.66 & 13.49 & I & 0.00011 & $6.25\times10^{-2}$ 
& $9.68\times 10^{-2}$ & 3.42 & 3.87 & 0.182 & 0.818 \\ 
B$1951+32$ & 39.53 & 11.99 & I & 0.00035 & $6.25\times10^{-2}$ 
& $1.62\times 10^{-4}$ & 2.97 & 1.63 & 0.956 & 0.044 \\ 
B$1046-58$ & 123.65 & 12.84 & I & 0.00022 & $6.25\times10^{-2}$ 
& $8.07\times 10^{-3}$ & 3.15 & 2.24 & 0.696 & 0.304 \\
B$1259-63$ & 47.76 & 11.82 & I & 0.00049 & $6.25\times10^{-2}$ 
& $7.43\times 10^{-5}$ & 2.85 & 1.54 & 0.970 & 0.030 \\
J$0537-6906$ & 16.11 & 12.26 & I & 0.00015 & $6.25\times10^{-2}$ 
& $5.94\times10^{-4}$ & 3.30 & 1.86 & 0.916 & 0.084 \\
B$1823-13$ & 101.45 & 12.75 & I & 0.00023 & $6.25\times10^{-2}$ 
& $5.26\times 10^{-3}$ & 3.14 & 2.13 & 0.753 & 0.247 \\ 
B$1800-21$ & 133.61 & 12.93 & I & 0.00021 & $6.25\times10^{-2}$ 
& $1.20\times 10^{-2}$ & 3.18 & 2.36 & 0.635 & 0.365 \\ 
B$1929+10$ & 226.52 & 12.02 & II & 0.00042 & $6.25\times10^{-2}$ 
& $1.50\times 10^{-4}$ & 2.21 & 1.38 & 0.958 & 0.042 \\ 
B$0656+14$ & 384.89 & 12.97 & II & 0.00026 & $6.25\times10^{-2}$ 
& $1.34\times 10^{-2}$ & 2.70 & 2.09 & 0.615 & 0.385 \\ 
B$0540-69$ & 50.37 & 13.00 & I & 0.00011 & $6.25\times10^{-2}$ 
& $1.59\times 10^{-2}$ & 3.41 & 2.61 & 0.583 & 0.417 \\ 
J$1105-6107$ & 63.19 & 12.31 & I & 0.00031 & $6.25\times10^{-2}$ 
& $7.01\times10^{-4}$ & 3.03 & 1.77 & 0.908 & 0.092 \\
B$0950+08$ & 253.07 & 11.69 & II & 0.00024 & $6.25\times10^{-2}$ 
& $4.39\times 10^{-5}$ & 1.55 & 1.15 & 0.977 & 0.023 \\ 
B$1610-50$ & 231.60 & 13.33 & I & 0.00017 & $6.25\times10^{-2}$ 
& $5.83\times 10^{-2}$ & 3.25 & 3.20 & 0.295 & 0.705 \\ 
J$0538+2817$ & 143.16 & 12.16 & II & 0.00070 & $6.25\times10^{-2}$ 
& $3.56\times 10^{-4}$ & 2.65 & 1.58 & 0.935 & 0.065 \\ 
B$1055-52$ & 197.10 & 12.34 & II & 0.00050 & $6.25\times10^{-2}$ 
& $7.66\times 10^{-4}$ & 2.59 & 1.61 & 0.904 & 0.096 \\ 
B$0355+54$ & 156.38 & 12.22 & II & 0.00064 & $6.25\times10^{-2}$ 
& $4.65\times 10^{-4}$ & 2.65 & 1.59 & 0.925 & 0.075 \\ 
B$2334+61$ & 495.28 & 13.29 & II & 0.00020 & $6.25\times10^{-2}$ 
& $4.96\times 10^{-2}$ & 2.81 & 2.67 & 0.334 & 0.666 \\ 
B$0823+26$ & 530.66 & 12.29 & II & 0.00012 & $6.25\times10^{-2}$ 
& $7.41\times 10^{-4}$ & 1.59 & 1.23 & 0.906 & 0.094 \\ 
B$1853+01$ & 267.40 & 13.18 & II & 0.00038 & $6.25\times10^{-2}$ 
& $3.30\times 10^{-2}$ & 3.11 & 2.71 & 0.430 & 0.570 \\
\hline 
J$0437-4715$ & 5.75 & 9.06 & II & 0.0081 & $9.38\times 10^{-2}$ 
& $5.74\times 10^{-10}$ & 1.49 & 1.05 & 0.9999 & 0.0001 \\ 
B$1937+21$ & 1.55 & 8.90 & II & 0.0415 & $9.38\times10^{-2}$ 
& $7.22\times10^{-10}$ & 2.50 & 1.17 & 0.9999 & 0.0001 \\
B$1821-24$ & 3.05 & 9.65 & II & 0.0216 & $9.38\times 10^{-2}$ 
& $2.46\times 10^{-8}$ & 2.74 & 1.23 & 0.9996 & 0.0004 \\ 
J$2124-3358$ & 4.93 & 8.67 & II & 0.0070 & $9.38\times 10^{-2}$ 
& $3.87\times 10^{-11}$ & 1.05 & 1.01 & 1.0000 & 0.0000 \\ 
B$1957+20$ & 1.60 & 8.52 & II & 0.0367 & $9.38\times 10^{-2}$ 
& $9.66\times 10^{-11}$ & 2.05 & 1.11 & 1.0000 & 0.0000 \\
J$1024-0719$ & 5.18 & 8.79 & II & 0.0074 & $9.38\times 10^{-2}$ 
& $9.26\times 10^{-11}$ & 1.19 & 1.02 & 1.0000 & 0.0000 \\ 
J$1744-1134$ & 4.07 & 8.58 & II & 0.0091 & $9.38\times 10^{-2}$ 
& $3.13\times 10^{-11}$ & 1.14 & 1.01 & 1.0000 & 0.0000 \\ 
J$1012+5307$ & 5.25 & 8.74 & II & 0.0068 & $9.38\times 10^{-2}$ 
& $5.83\times 10^{-11}$ & 1.09 & 1.01 & 1.0000 & 0.0000 \\
J$0218+4232$ & 2.32 & 8.94 & II & 0.0263 & $9.38\times 10^{-2}$ 
& $7.42\times 10^{-10}$ & 2.21 & 1.14 & 0.9999 & 0.0001 \\ 
J$0751+1807$ & 3.47 & 8.53 & II & 0.0116 & $9.38\times 10^{-2}$ 
& $3.17\times 10^{-11}$ & 1.26 & 1.03 & 1.0000 & 0.0000 \\
 
\hline 
\end{tabular} 
 
\end{table} 
 
\begin{table} 
\caption{X-ray luminosities observed and expected in the ROSAT 
 band and ASCA bands. ROSAT data follow BT97, and ASCA data follow S98.
 The middle 6 columns are for the ROSAT band, and the right 3 columns
 are for the ASCA band. `o' denotes
 observed value and `m' denotes model prediction. the superscript `tot'
 indicates total luminosity detected, `pul' indicates the luminosity of
 the pulsed portion, `nth' denotes non-thermal, and `th1' and `th2' 
 denote thermal emission of full surface and polar cap,
 respectively.}
 
\begin{tabular}{lr|clcccc|crl} 
\hline 
\hline 
Pulsars & $L_{\rm sd}$ & & & ROSAT & band & & & & ASCA & band  \\
\hline
& & $L_x^{tot}$(o) & $L_x^{pul}$(o) &
$L_x^{th1}$(m) & $L_x^{th2}$(m) & $L_x^{nth}$(m) & 
$L_x^{tot}$(m) & $L_x^{tot}$(o) & $L_x^{pul}$(o) & $L_x$(m) \\
\hline 
B$0531+21$ & 38.65 & 35.98 & 35.85 & 32.49 & $\lesssim 31.55$ 
& $ < 34.25$ & $ < 34.26$ & 37.02 & 35.83 & $ < 34.83$ \\ 
B$0833-45$ & 36.84 & 32.70 & 31.7 & 32.96 & $\lesssim 30.98$ 
& $ < 33.35$ & $ < 33.50$ & 33.28 & $<31.83$ & $<33.96$ \\
J$1811-1926$ & 37.97 & & & 32.45 & $\lesssim 31.24$ 
& $ < 34.23$ & $ < 34.24$ & & & $<34.63$ \\
J$1617-5055$ & 37.22 & & & 32.17 & $\lesssim 31.11$ 
& $ < 33.73$ & $ < 33.74$ & & & $<34.21$ \\
B$0633+17$ & 34.51 & 31.10 & 30.62 & 31.24 & $\lesssim 30.18$ 
& $ < 31.72$ & $ < 31.86$ & 29.79 & 29.56 & $\lesssim 32.33$ \\ 
B$1706-44$ & 36.53 & 33.15 & & 32.03 & $\lesssim 30.89$ 
& $ < 33.46$ & $ < 33.48$ & 32.83 & $< 32.73$ & $ < 33.91$ \\
B$1509-58$ & 37.25 & 34.29 & 34.10 & 32.45 & $\lesssim 30.88$ 
& $ < 33.74$ & $ < 33.76$ & 34.60 & 34.80 & $< 34.34$ \\ 
B$1951+32$ & 36.57 & 33.44 & & 31.72 & $\lesssim 31.19$ & $\lesssim 
32.18$ & $\lesssim 32.34$ & 33.79 & $<32.93$ & $\lesssim 32.78$ \\ 
B$1046-58$ & 36.30 & $\leq 32.11$ & & 32.01 & $\lesssim 30.80$ & $ <33.34$ 
& $ < 33.36$ & 32.96 & $< 32.78$ & $< 33.80$\\
B$1259-63$ & 35.92 & 32.95 & & 31.00 & $\lesssim 31.04$ 
& $\lesssim 31.92$ & $\lesssim 32.02$ & & & $\lesssim 32.53$ \\ 
J$0537-6906$ & 38.68 & & & 32.25 & $\lesssim 31.78$
& $ < 34.24$ & $ < 34.25$ & & & $\lesssim 34.45$ \\ 
B$1823-13$ & 36.45 & 33.39 & & 32.00 & $\lesssim 30.88$ 
& $ < 33.45$ & $ < 33.46$ & & & $\lesssim 33.83$ \\ 
B$1800-21$ & 36.35 & 33.06 & & 32.05 & $\lesssim 30.78$ 
& $ < 33.31$ & $ < 33.33$ & & & $ < 33.80$ \\ 
B$1929+10$ & 33.60 & 30.00 & 29.5 & 29.07 & $\lesssim 29.72$ 
& $< 31.63$ & $< 31.64$ & 30.08 & $<29.92$ 
& $< 31.70$ \\ 
B$0656+14$ & 34.58 & 32.98 & 32.15 & 31.86 & $\lesssim 29.95$ 
& $ < 32.24$ & $ < 32.39$ & & $<31.23$ & $<32.83$ \\ 
B$0540-69$ & 38.17 & 36.21 & 36.1 & 32.44 & $\lesssim 31.35$ 
& $ < 33.98$ & $ < 33.99$ & 36.92 & 36.11 & $<34.58$ \\ 
J$1105-6107$ & 36.39 & & & 31.81 & $\lesssim 31.02$
& $ < 33.34$ & $ < 33.35$ & & & $\lesssim 33.41$ \\ 
B$0950+08$ & 32.75 & 29.35 & & 27.56 & $\lesssim 28.88$ 
& $< 30.28$ & $< 30.30$ & 29.00 & $<29.00$ 
& $< 30.84$ \\ 
B$1610-50$ & 36.19 & & & 32.18 & $\lesssim 30.59$ 
& $<33.59$ & $<33.61$ & 34.36 & $<34.36$ & $< 34.12$ \\ 
J$0538+2817$ & 34.69 & 32.74 & & 30.46 & $\lesssim 30.55$ 
& $ <31.81$ & $< 31.85$ & & & $\lesssim 32.42$ \\
B$1055-52$ & 34.48 & 33.42 & 32.57 & 31.86 & $\lesssim 30.26$
& $ <31.49$ & $ < 32.03$ & 30.45 & $<30.43$ & $\lesssim 32.10$ \\ 
B$0355+54$ & 34.66 & 31.96 & & 30.54 & $\lesssim 30.48$ 
& $ <31.82$ & $ < 31.86$ & & & $\lesssim 32.43$ \\ 
B$2334+61$ & 34.80 & 31.86 & & 31.89 & $\lesssim 29.93$ 
& $ < 32.15$ & $ < 32.34$ & & & $ < 32.75$ \\ 
B$0823+26$ & 32.66 & 29.83 & & 28.67 & $\lesssim 28.45$ 
& $ < 30.74$ & $ < 30.75$ & & & $\lesssim 31.32$ \\ 
B$1853+01$ & 35.63 & & & 32.01 & $\lesssim 30.69$ 
& $<32.83$ & $<32.89$ & 33.15 & $<33.15$ & $<33.33$ \\
\hline 
J$0437-4715$ & 34.07 & 30.86 & 30.3 & 23.64 & $\lesssim 31.30$ 
& $\sim 29.82$ & $\lesssim 31.31$ & 30.08 & $<30.08$ & $\sim 30.06$ \\ 
B$1937+21$ & 36.03 & $\leq 32.10$ & & 25.27 & $\lesssim 33.13$
& $\sim 31.22$ & $\lesssim 33.13$ & & & $\sim 31.46$ \\
B$1821-24$ & 36.35 & 33.24 & & 27.09 & $\lesssim 32.93$ 
& $\sim 31.54$ & $\lesssim 32.95$ & 33.81 & 32.97 & $\sim 31.77$ \\ 
J$2124-3358$ & 33.55 & 30.35 & 29.8 & 22.33 & $\lesssim 30.99$ 
& $\sim 29.49$ & $\lesssim 31.01$ & & & $\sim 29.74$ \\ 
B$1957+20$ & 35.20 & 31.93 & & 23.70 & $\lesssim 32.66$ 
& $\sim 30.78$ & $\lesssim 32.67$ & & & $\sim 31.03$ \\ 
J$1024-0719$ & 33.71 & 29.48 & & 22.73 & $\lesssim 31.09$ 
& $\sim 29.59$ & $\lesssim 31.10$ & & & $\sim 29.83$ \\ 
J$1744-1134$ & 33.70 & 29.30 & & 22.30 & $\lesssim 31.20$ 
& $\sim 29.54$ & $\lesssim 31.21$ & & & $\sim 29.78$ \\
J$1012+5307$ & 33.58 & 30.20 & & 22.50 & $\lesssim 30.99$ 
& $\sim 29.52$ & $\lesssim 31.00$ & & & $\sim 29.77$ \\
J$0218+4232$ & 35.40 & 32.75 & & 24.72 & $\lesssim 32.57$ 
& $\sim 30.96$ & $\lesssim 32.58$ & & & $\sim 31.20$ \\
J$0751+1807$ & 33.88 & 31.60 & & 22.37 & $\lesssim 31.41$ 
& $\sim 29.60$ & $\lesssim 31.42$ & & & $\sim 29.85$ \\
\hline 
\end{tabular} 
 
\end{table}

\end{document}